\documentclass[sigconf]{acmart}

\usepackage{tikz}

\usepackage{filecontents}
\usepackage{multirow}
\usepackage{textcomp}
\usepackage{balance}
\usepackage{tabularx}
\usepackage{caption}
\usepackage{subcaption}
\usepackage{graphicx}
\usepackage{bbm}
\usepackage{enumitem}
\usepackage{subfiles}
\usepackage{xurl}
\usepackage[para,online,flushleft]{threeparttable}
\usepackage[export]{adjustbox}
\usepackage{natbib}
\usepackage{epstopdf}
\usepackage{algorithmic}
\usepackage{hyperref}
\usepackage{hyphenat}
\usepackage[]{algorithm2e} 
\usepackage{fancyhdr,amsmath}
\usepackage{booktabs,subcaption,amsfonts,dcolumn}
\usepackage{xcolor}
\usepackage{color, colortbl}
\usepackage{textcomp}
\usepackage{listings}
\usepackage{array}

\newcommand{\COMMENT}{\textcolor{red}}

\colorlet{punct}{red!60!black}
\definecolor{background}{HTML}{EEEEEE}
\definecolor{delim}{RGB}{20,105,176}
\colorlet{numb}{magenta!60!black}

\lstdefinelanguage{json}{
    basicstyle=\normalfont\ttfamily,
    numbers=left,
    numberstyle=\scriptsize,
    stepnumber=1,
    numbersep=8pt,
    showstringspaces=false,
    breaklines=true,
    frame=lines,
    backgroundcolor=\color{background},
    literate=
     *{0}{{{\color{numb}0}}}{1}
      {1}{{{\color{numb}1}}}{1}
      {2}{{{\color{numb}2}}}{1}
      {3}{{{\color{numb}3}}}{1}
      {4}{{{\color{numb}4}}}{1}
      {5}{{{\color{numb}5}}}{1}
      {6}{{{\color{numb}6}}}{1}
      {7}{{{\color{numb}7}}}{1}
      {8}{{{\color{numb}8}}}{1}
      {9}{{{\color{numb}9}}}{1}
      {:}{{{\color{punct}{:}}}}{1}
      {,}{{{\color{punct}{,}}}}{1}
      {\{}{{{\color{delim}{\{}}}}{1}
      {\}}{{{\color{delim}{\}}}}}{1}
      {[}{{{\color{delim}{[}}}}{1}
      {]}{{{\color{delim}{]}}}}{1},
}

\newcolumntype{H}{>{\setbox0=\hbox\bgroup}c<{\egroup}@{}}

\AtBeginDocument{%
  \providecommand\BibTeX{{%
    \normalfont B\kern-0.5em{\scshape i\kern-0.25em b}\kern-0.8em\TeX}}}

\setcopyright{acmcopyright}
\copyrightyear{2022}
\acmYear{2022}
\acmDOI{10.1145/1122445.1122456}

\acmConference[ICCPS' 22]{ICCPS '22: ACM/IEEE International Conference on Cyber-Physical Systems}{May 4-6, 2022}{Milan, Italy}
\acmBooktitle{ICCPS '22: ACM/IEEE International Conference on Cyber-Physical Systems, May 4-6, 2022, Milan, Italy}
\acmPrice{15.00}
\acmISBN{978-1-4503-XXXX-X/18/06}

\renewcommand\footnotetextcopyrightpermission[1]{}

\begin{document}

\title[Infrastructure-free, Deep Learned Urban Noise Monitoring at $\sim$100mW]{Infrastructure-free, Deep Learned Urban Noise Monitoring \\ at $\sim$100mW}


\author{Jihoon Yun}
\affiliation{
  \institution{The Ohio State University}
  \country{}}
\email{yun.131@osu.edu}

\author{Sangeeta Srivastava}
\affiliation{
  \institution{The Ohio State University}
  \country{}}
\email{srivastava.206@osu.edu}

\author{Dhrubojyoti Roy}
\affiliation{
  \institution{The Ohio State University}
  \country{}}
\email{roy.174@osu.edu}

\author{Nathan Stohs}
\affiliation{%
  \institution{The Samraksh Company}
  \country{}}
\email{nathan.stohs@samraksh.com}

\author{Charlie Mydlarz}
\affiliation{%
  \institution{New York University}
  \country{}}
\email{cmydlarz@nyu.edu}

\author{Mahin Salman}
\affiliation{%
  \institution{New York University}
  \country{}}
\email{ms6617@nyu.edu}

\author{Bea Steers}
\affiliation{%
  \institution{New York University}
  \country{}}
\email{bsteers@nyu.edu}

\author{Juan Pablo Bello}
\affiliation{%
  \institution{New York University}
  \country{}}
\email{jpbello@nyu.edu}

\author{Anish Arora}
\affiliation{%
  \institution{The Ohio State University}
  \country{}}
\email{arora.9@osu.edu}

\renewcommand{\shortauthors}{J. Yun, S. Srivastava, D. Roy, N. Stohs, C. Mydlarz, M. Salman, B. Steers, J. P. Bello, A. Arora}
%

\begin{abstract}
The Sounds of New York City (SONYC) wireless sensor network (WSN) has been fielded in Manhattan and Brooklyn over the past five years, as part of a larger human-in-the-loop cyber-physical control system for monitoring, analyzing, and mitigating urban noise pollution. We describe the evolution of the 2-tier SONYC WSN from an acoustic data collection fabric into a 3-tier in situ noise complaint monitoring WSN, and its current evaluation. The added tier consists of long range (LoRa), multi-hop networks of a new low-power acoustic mote, MKII (``Mach 2''), that we have designed and fabricated. MKII motes are notable in three ways: First, they advance machine learning capability at mote-scale in this application domain by introducing a real-time Convolutional Neural Network (CNN) based embedding model that is competitive with alternatives while also requiring 10$\times$ lesser training data and {\small $\sim$}2 orders of magnitude fewer runtime resources. Second, they are conveniently deployed relatively far from higher-tier base station nodes without assuming power or network infrastructure support at operationally relevant sites (such as construction zones), yielding a relatively low-cost solution. 
And third, their networking is frequency agile, unlike conventional LoRa networks: it tolerates in a distributed, self-stabilizing way the variable external interference and link fading in the cluttered 902-928MHz ISM band urban environment by dynamically choosing good frequencies using an efficient new method that combines passive and active measurements. 

\end{abstract}


\maketitle

\section{Introduction}

%
%
%

Sounds of New York City \cite{bello2019sonyc} is a large-scale WSN deployed at operationally relevant locations in Manhattan, Brooklyn and Queens to facilitate monitoring and mitigation of urban noise complaints---a true health hazard in megacities like New York City that impairs the quality of life of its denizens \cite{basner2014auditory,bronzaft2010noise,fritschi2012introduction,isling2004}. Since its inception in 2016, the deployed system has collected audio recordings and sound pressure level (SPL) data using a network of 55 microphone-equipped Raspberry Pi 2B-based MKI  (``Mach 1'') devices \cite{mydlarz2019life}. This Tier 1 network is managed by a Tier 0 private cloud server infrastructure. The data has enabled offline analysis across the interdisciplinary domains of machine listening and citizen science, and also online use by a key partner, the New York City Department of Environmental Protection (DEP), to guide planning of its inspection activities. However, the current system has three limitations:

\begin{figure}[t]
	\centering
	\includegraphics[width=0.72\linewidth]{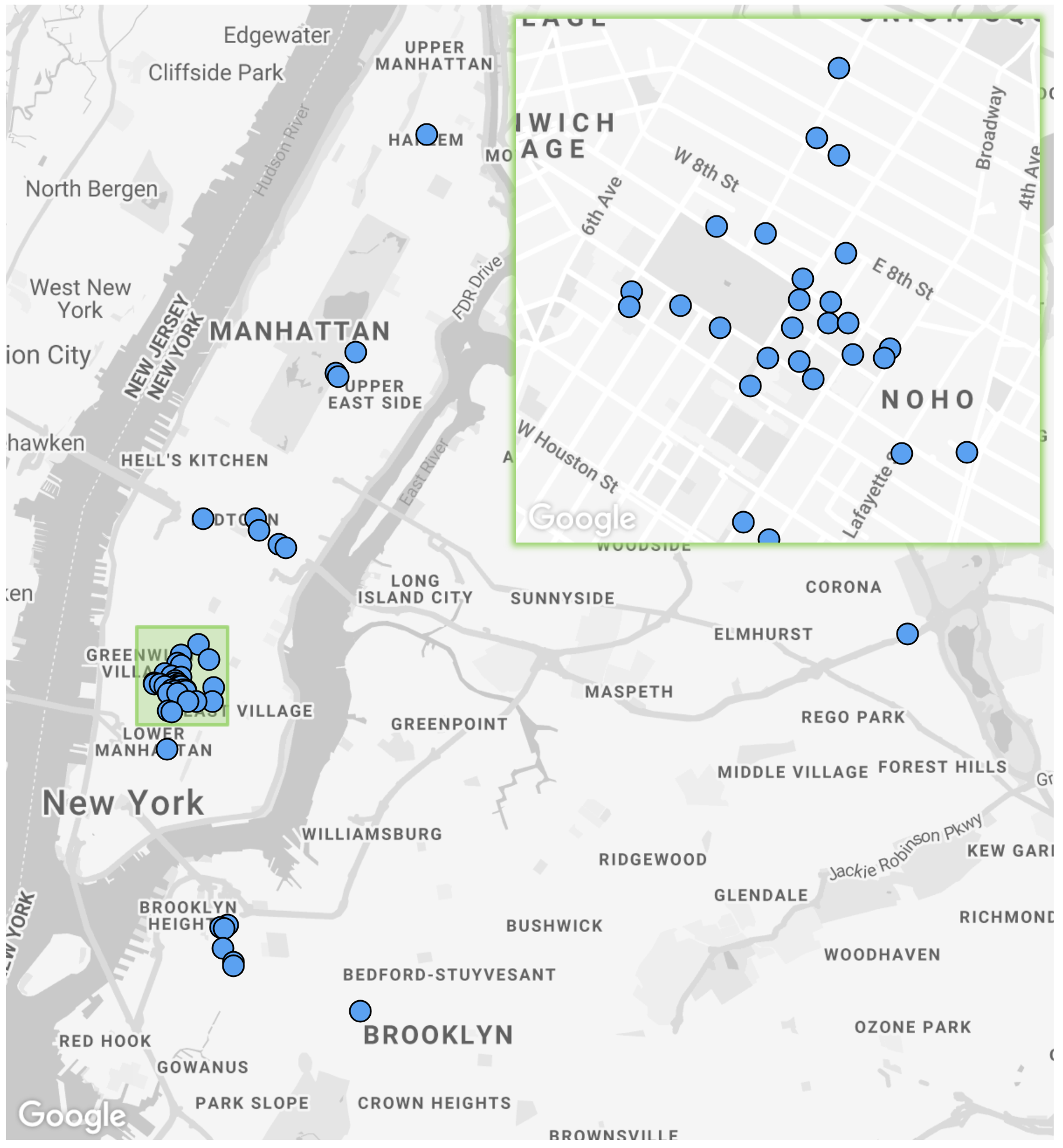}
	\caption{The current SONYC deployment is infrastructure-aided and hence can only cover a small fraction of NYC (Image courtesy: \cite{mydlarz2019life})}
	\label{fig:sonyc_phase1_deployment}
\end{figure}

\begin{enumerate}[label=(\roman*),topsep=0pt,itemsep=-1ex,partopsep=1ex,parsep=1ex]
    \item SONYC currently does not incorporate real-time classification of noise sources. This feature is desirable as the 311 line for registering noise complaints in NYC receives {\small $\sim$}1300 calls per day on average (as analyzed with 2019 data from the 311 open dataset \cite{311dataset}), with a large number caused by construction practices, that are often not timely enough to be actionable. The converse problem also exists: a fraction of these complaints turn out to be false alarms upon investigation, thereby wasting DEP and Police Department (NYPD) resources. 
   \item MKI devices require infrastructure power support, which adds significant procedural complexity, time, and cost to their deployment. 
	\item The current deployment uses existing city Wi-Fi infrastructure (Figure \ref{fig:sonyc_phase1_deployment}), which limits the achievable area that can be covered in NYC (to only {\small $\sim$}3\% of the city if leveraging LinkNYC Wi-Fi, as analyzed with its open data set \cite{linknyclocation}). 
\end{enumerate}

\begin{figure}[t]
	\centering
	\includegraphics[width=0.9\linewidth]{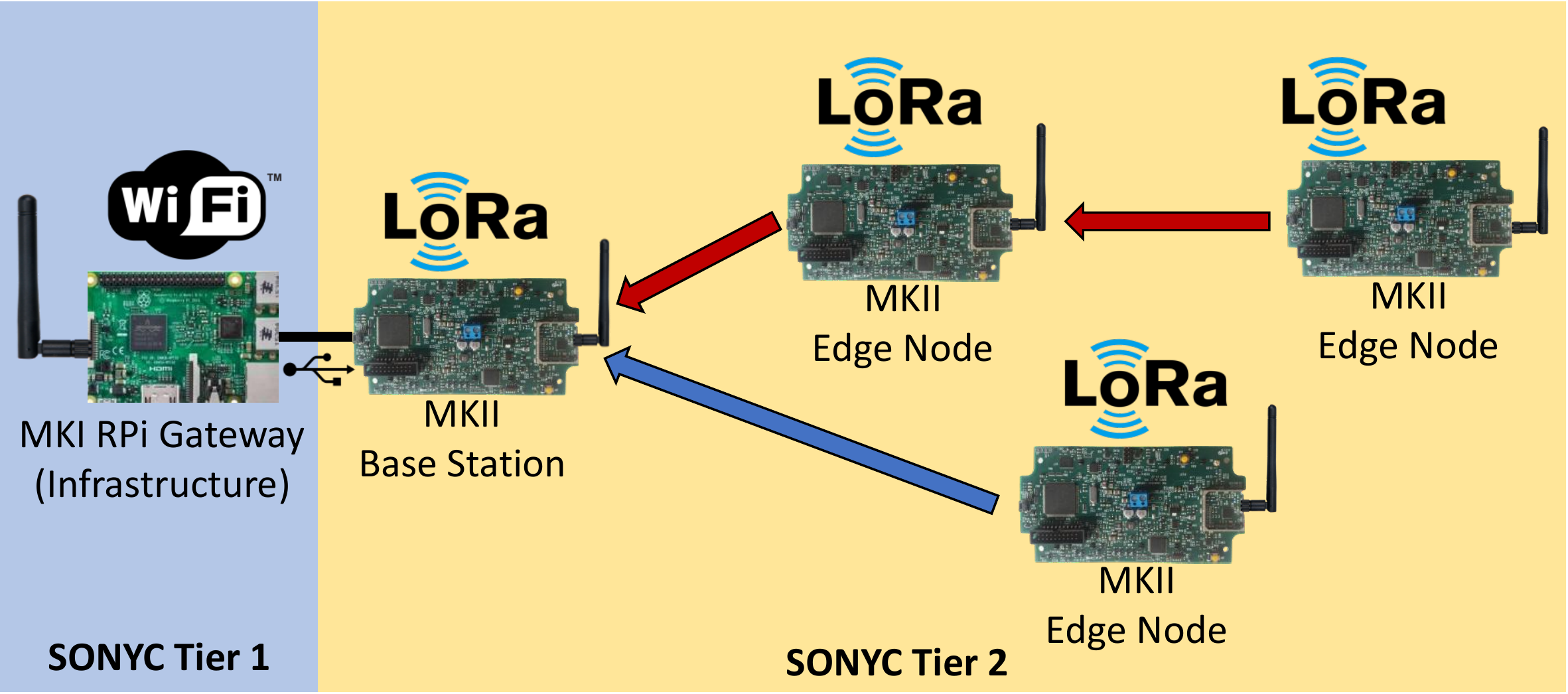}
    \vspace{-1mm}
	\caption{SONYC Tier 2 is an 
	infrastructure-agnostic extension to the current data collection fabric that offers improved coverage with cheap installation as well as real-time noise complaint discrimination capabilities in situ}
	\label{fig:sonyc_phase_2}
	\vspace{-3.5mm}
\end{figure}

We describe the enhancement of SONYC from a data collection system to a software-defined, infrastructure-free sensing fabric that redresses the three current limitations of SONYC (see Figure\ref{fig:sonyc_phase_2}).
\vspace{1mm}

\noindent
{\bf Contributions of this paper}.~~~ At the heart of the MKII WSN is a powerful mote-scale machine listening CNN, SONYC-L\textsuperscript{3}, which to the best of our knowledge is \textit{the first real-time mote embedding model for urban sound classification}. Its performance is competitive with, and often better than, that of other embedding models in literature, while also requiring \textit{{\small $\sim$}2 orders of magnitude fewer resources}. This novel embedding architecture tackles the high activation memory of the reference CNN, L\textsuperscript{3}-Net \cite{arandjelovic2017look}, through input resolution reduction and aggressive filter dropping. The model can be efficiently trained with Specialized Embedding Approximation \cite{srivastava2021specialized}, a variant of knowledge distillation that offers superior compression while requiring up to \textit{an order of magnitude lesser training data}. The overall runtime power consumption of the networked mote app in classification mode is \textit{only 107~mW} (the mote app also offers a continuous SPL meter mode, which offers a substantially lower cost alternative than existing commercial noise meters such as those used by airports in NYC).


To support infrastructure-free MKII operation, the acoustic inference module is deployed on a custom-designed low-power mote that we developed, which is based on an ARM Cortex-M7 and a Cortex-M3 processor and interfaced to a low-power audio front-end. MKII is powered by a small, 5W solar harvester with sufficient battery capacity and a software-defined battery managed subsystem that yields only rare outages. Its applications are supported by the eMote, a derivative of the .NET Micro Framework runtime. Notably, eMote has been refined to support \textit {high-level programmability of components that operate with low-jitter and low-power}, including components for ML, wireless networking, and management.

The small form factor and self-powered design of MKII yields convenient, low-cost deployment. MKII devices can be emplaced on available natural or man-made structures at significant distances from available Wi-Fi infrastructure.  To this end, MKII devices support a self-healing, low-power, multi-hop ``long range'' (LoRa) wireless network where link ranges of at least 500m are readily supported despite the varied urban/wireless clutter conditions of SONYC deployments. The WSN is scheduled by an ultra low duty cycle MAC protocol, OMAC \cite{
configurationspace}. 

Notably, unlike conventional LoRa networks, our solution \textit {dynamically tolerates the high degree of external interference and link fading in the city without unduly decreasing data rate}, while being compliant to Federal Communications Commission (FCC) guidelines. It does so by dynamically selecting the frequency associated with links, via an efficient method that performs passive measurements and then selective active frequency measurements to select a common frequency that (near) optimizes network reliability. The limited capacity of LoRa links is accommodated by per-sensor aggregation of classifications and network measurements to reduce network traffic. A Collection Tree Protocol (CTP)-based routing mechanism relays the aggregate messages to the nearest MKII Base Station-MKI Gateway pair.  Together, the average power consumption of the radio and network components is below 15mW.

We have fabricated 100 MKII motes for deployment and have been progressively growing the Tier 2 network in downtown Brooklyn (as well as elsewhere in an airport monitoring and in-building setting; these are for robustness and other evaluations although those discussions are beyond the scope of this paper).  Concurrently, we have been testing and validating its network and application level performance over several months, in addition to collecting new data with the Tier 2 network.   


In sum, the SONYC Tier 2 system that we describe in this paper enables complex edge machine listening for noise sources in an infrastructure-agnostic manner. It allows coverage to be expanded to more operationally relevant locations, such as construction sites and airports. It is deployable in a relatively affordable and easy manner, 
while preserving the SONYC system capability of being managed via the cloud with limited effort. And its software-defined platform has allowed for repurposing its application, i.e., for other smart city contexts. Finally, we intend to open source a dockerized system that should work out-of-the box on compatible hardware, as well as its individual components: the OS, ML models, and training and quantization pipelines, that can be leveraged in other sensing applications as appropriate \footnote{\url{https://github.com/sonyc-project/SONYC-MKII}}.

\section{System Overview and Preliminaries}

\subsection{System Overview}
\begin{figure}[t]
	\centering
	\includegraphics[width=1.25\linewidth, trim = 75 0 5 0, clip]{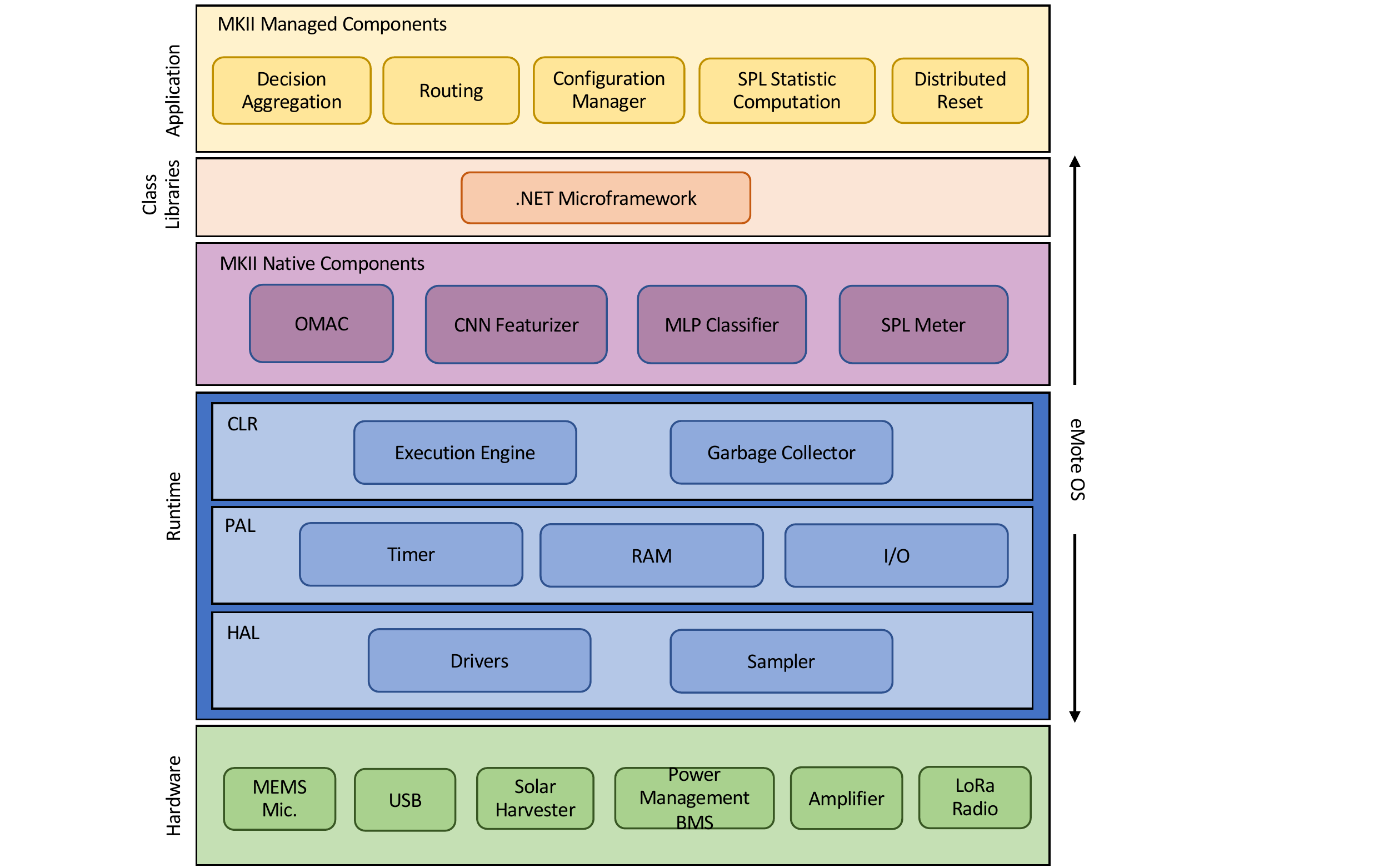}
	\vspace{-2mm}
	\caption{Overview of the SONYC MKII system, where user apps run in a virtual machine or as native components, supported by the eMote runtime}
	\label{fig:sysmodel}
	\vspace{-4.5mm}
\end{figure}

Figure \ref{fig:sysmodel} shows a simplified overview of the MKII hardware-software system that we developed for SONYC. It has four main hardware components: For sensing, a digital I2S MEMS microphone based acoustic front end. For RF communication, a SX1276 LoRa radio chip with external amplifier. For power, a solar harvester, 4 lithium-titanate (LTO) cells, and a Cortex-M3 based microcontroller (STM32 F103) for a power management subsystem. For computing, a Cortex-M7 microcontroller (STM32H753) along with an external 1 MB RAM and 16 MB QSPI Flash.

We ported eMote \cite{emote}, a runtime environment for mote-scale device to the MKII. eMote is a substantially stripped down version of the open source .Net Micro Framework \cite{netmf} that enables low-power, low-jitter, near real time computing and wireless networking. Its common language runtime (CLR) supports the execution of managed components programmed in C\# in a virtual machine, as well as direct execution of components programmed in C++.


The SONYC application consists of five managed and four native components. The managed components include:  A Decision Aggregation app that aggregates classifier decisions over a time window and periodically communicates them to the base station.  A SPL Meter app that computes the sound pressure statistics and likewise communicates periodically. A Router that maintains the node's neighborhood and selects paths to the base station using the CTP protocol \cite{gnawali2009collection}. A Configuration Manager component that handles commands from the base station to change application parameters (such as aggregation rate, sampling rates, etc.). And a fault tolerant Distributed Reset \cite{arora1994distributed} component that self-heals the routing network in the presence of link or node dynamics, and is used to program the flood to reliably command all nodes. The native components include an acoustic pre-processor, an embedding CNN model, a downstream noise source classifier, and a SPL meter.

\vspace*{-1mm}
\subsubsection{Integration with Tier 1}
Tier 2 integrates into the existing SONYC infrastructure, 
utilizing SONYC's existing data transfer, storage, and visualization systems. A base station MKII node is tethered via USB to a MKI node and communicates to it over serial. A background service on the MKI handles the MKII message passing and provides a simple JSON interface over a UDP socket. Information received by the MKI from the MKII network is uploaded to the existing Tier 0 Elasticsearch database in a data ingestion cluster and is made available via sensor dashboards. The UDP socket also makes available an interface to propagate configuration updates to the MKII base station and edge node using JSON configurations, as described above.


\vspace*{-1mm}
\subsection{Reference Embedding Models: L$^3$-Net and EdgeL$^3$}
\label{subsec_avc}
Owing to the lack of sufficient labeled data in the urban sound classification domains, researchers opt for \textit{transfer learning} to achieve generalizability. With this approach, one can train a model through unsupervised or self-supervised methods on large amount of unlabeled ``upstream'' audio data and subsequently use it as a robust featurizer on various domain-specific ``downstream'' contexts. Look, Listen, Learn (L\textsuperscript{3}-Net) \cite{arandjelovic2017look} is one such self-supervised model trained to learn representations or embeddings via the audio-visual correspondence (AVC) task. This auxiliary task aims to predict whether a 1s audio segment and a single video image frame come from the same video and also overlap in time. The learned audio embeddings can then be used in various \textit{downstream} scenarios such as acoustic event detection \cite{cramer2019}, making it a suitable model to be adopted for our application. However, its formidable storage (18 MB) and activation (12 MB) memory requirements make it quite challenging to be implemented at the scale of a MKII mote.

Previous work, such as EdgeL\textsuperscript{3} \cite{kumari2019edgel}, has attempted to solve the storage problem through magnitude-based sparsification, yielding the first edge reference model for urban machine listening with sensing performance comparable to L\textsuperscript{3}-Net with $>$95\% sparsification. In particular, this work has demonstrated how aggressive sparsification, in conjunction with post hoc fine-tuning or knowledge distillation, can successfully alleviate the storage problem without compromising downstream sensing quality. However, EdgeL\textsuperscript{3} has failed to address the activation memory problem: activations of first two convolution layers of EdgeL\textsuperscript{3} require a dynamic memory of {\small$\sim$}12 MB, making the model infeasible for Cortex-M7 devices with only 1 MB of SRAM. 


\section{SONYC-L\textsuperscript{3}: Downscaling L$^3$-Net to Motes}

For real-time implementation on the MKII Cortex-M7, we use a different strategy to produce a smaller variant of  L\textsuperscript{3}-Net Audio. We leverage coarse-grained input processing, coupled with reducing model \textit{width} by halving the number of convolution filters at each layer, to reduce the activation memory by \textit{more than 1.2 orders of magnitude}. Further, with 8-bit integer quantization, the model has a dynamic memory footprint of only $\sim$120 KB and runs in $\sim$800 ms on each second of audio input, achieving truly real-time operation on the edge. We refer to this architecture as SONYC-L\textsuperscript{3} and evaluate it on the SONYC-UST downstream dataset (Section \ref{sec_sonyc_ust}).

\vspace{-2mm}
\subsection{The SONYC-UST Dataset}
\label{sec_sonyc_ust}
Since its inception in 2016, the SONYC sensor network has continuously collected urban audio data (We note that SONYC data collection and system development has received exemption from IRB approval, based on its data collection and processing methodology). Through subsequent crowdsourcing efforts on the Zooniverse \cite{zooniverse} platform, a fraction of this data has been annotated and recently released as the \textit{SONYC Urban Sound Tagging} (SONYC-UST) dataset \cite{cartwright2019sonyc}. This is a collection of 3068 10-second clips that were manually annotated for the presence or absence of a number of 8 sound events of interest. The dataset exhibit a class imbalance, with \textit{engine} being the most prominent class with 50\% of the data, while \textit{human-voice} and \textit{dog} have only 5\% and 6\% of the data, respectively. 


SONYC-UST is a \textit{multi-label} task, where the presence or absence of each label in the 10-second clips are mutually non-exclusive. As advised in \cite{cartwright2019sonyc}, we use macro- and micro-averaged areas under the precision-recall curve (AUPRC) as the primary evaluation metrics for this dataset, along with an additional (secondary) metric of micro-averaged F1 scores at a threshold of 0.5. We also report the class-specific AUPRCs of each class for a finer-grained assessment of classifier performance.

\begin{figure}[t]
    \centering
	\begin{subfigure}[t]{0.48\textwidth}
	\centering
		\begin{subfigure}[t]{0.5\textwidth}
		    \centering
			\includegraphics[width=\textwidth]{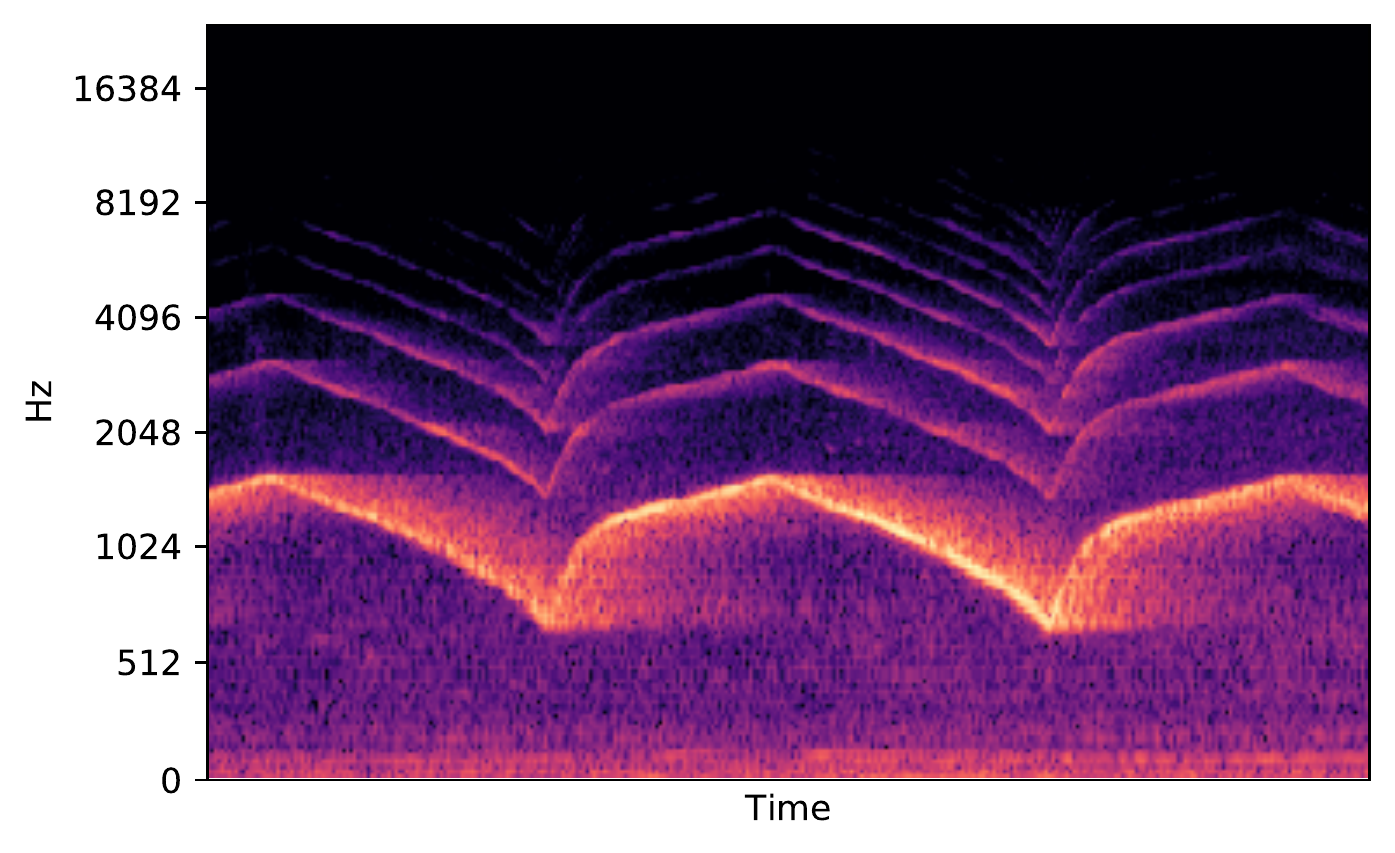}
			\caption*{(i) alert-signal: L\textsuperscript{3}}
		\end{subfigure}%
		~
		\begin{subfigure}[t]{0.5\textwidth}
		    \centering
			\includegraphics[width=\textwidth]{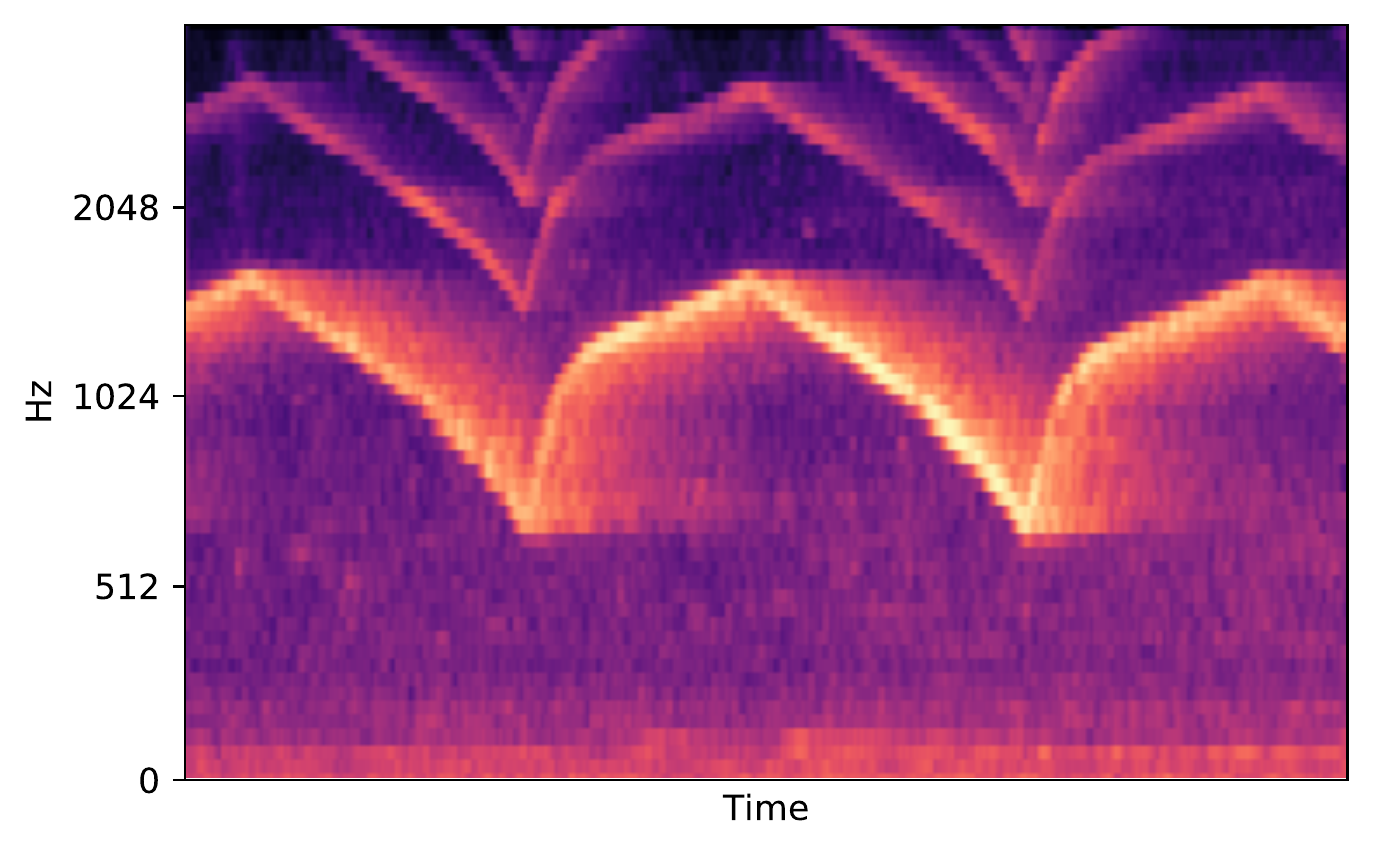}
			\caption*{(ii) alert-signal: SONYC-L\textsuperscript{3}}
		\end{subfigure}%
	\end{subfigure}
	
	\begin{subfigure}[b]{0.48\textwidth}
	    \centering
		\begin{subfigure}[b]{0.5\textwidth}
		    \centering
			\includegraphics[width=\textwidth]{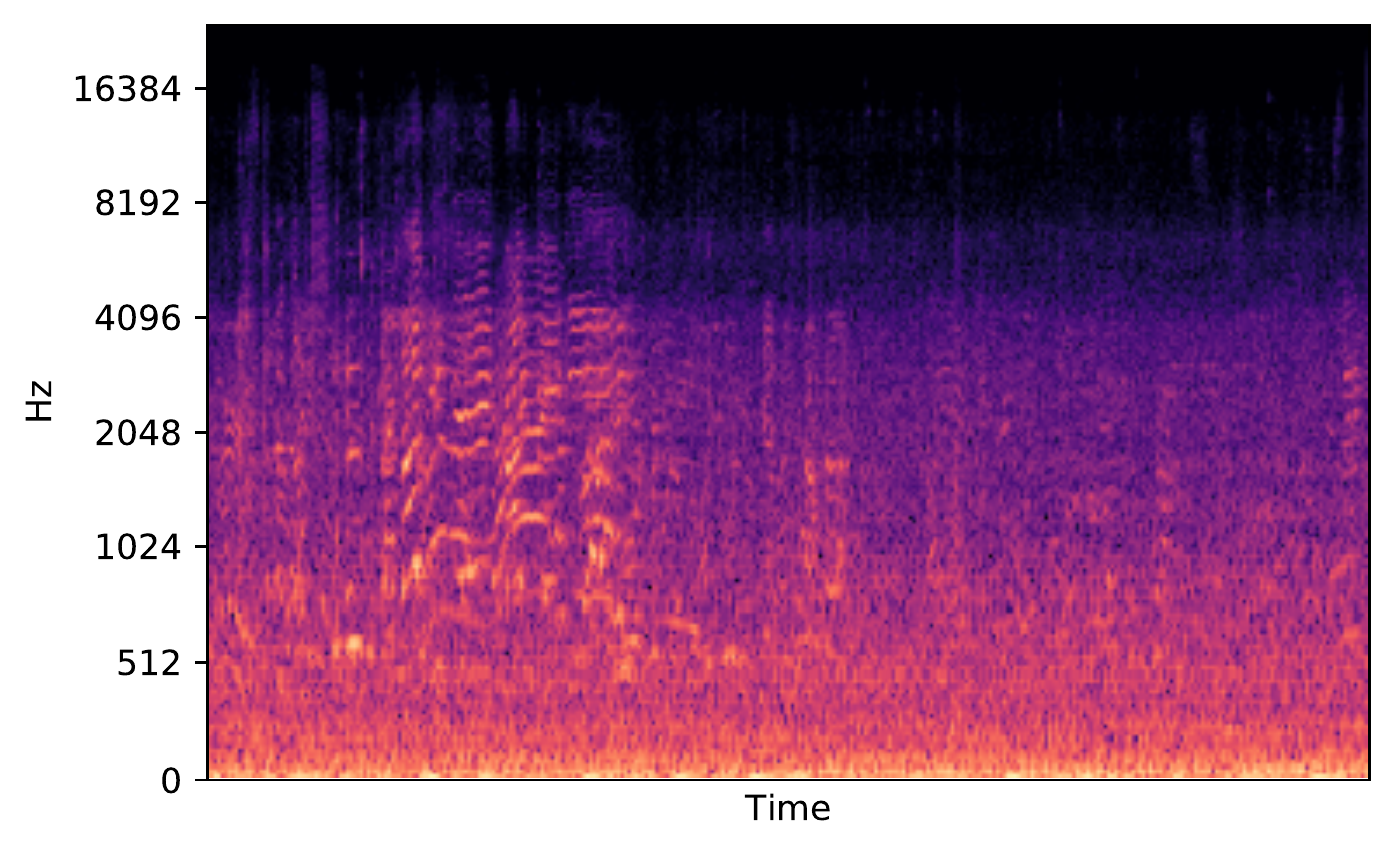}
			\caption*{(iii) human-voice: L\textsuperscript{3}}
		\end{subfigure}%
        ~
		\begin{subfigure}[b]{0.5\textwidth}
		    \centering
			\includegraphics[width=\textwidth]{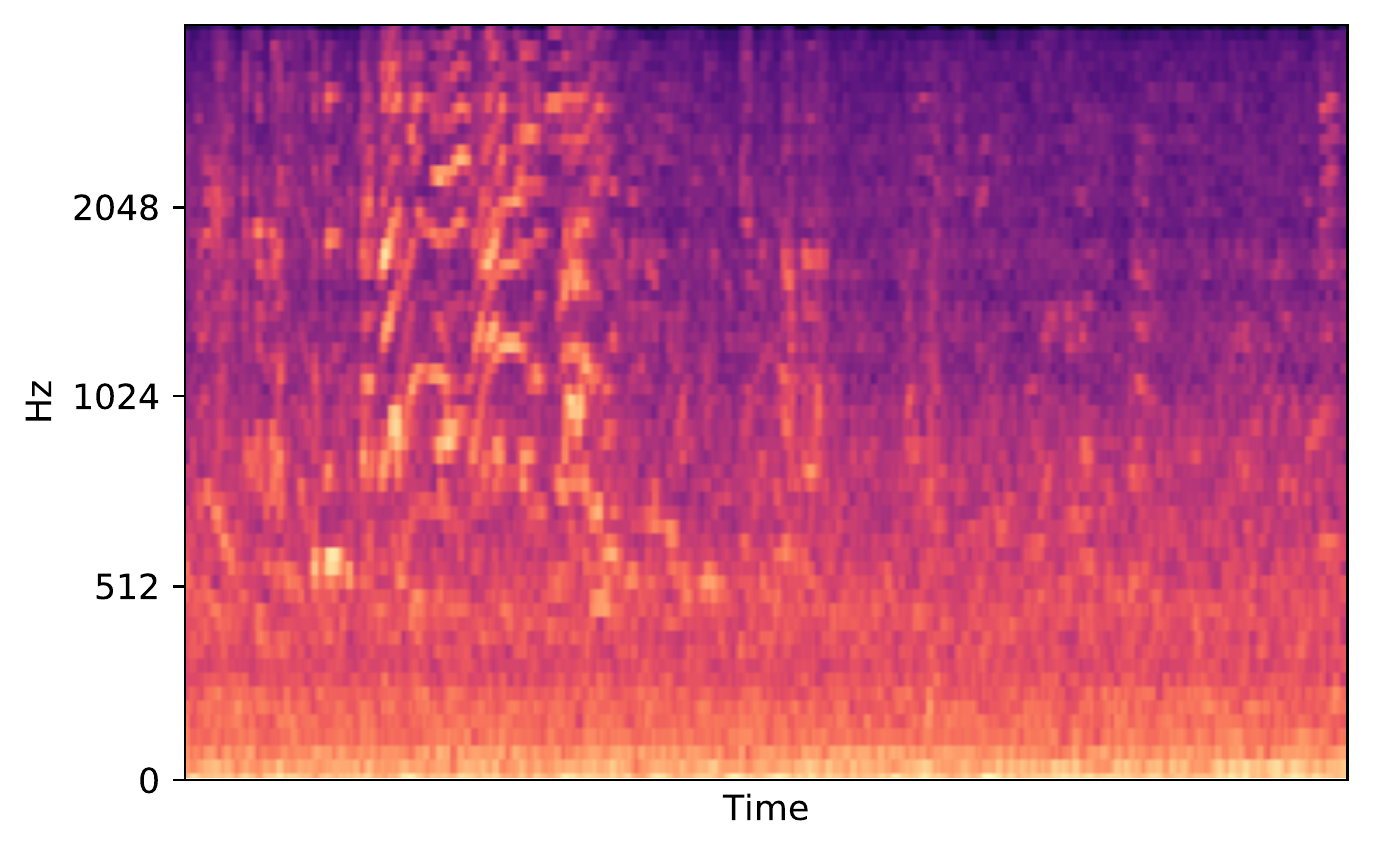}
			\caption*{(iv) human-voice: SONYC-L\textsuperscript{3}}
		\end{subfigure}%
	\end{subfigure}
	\caption{Melspectrograms of representative noise sources in fine-grained (L\textsuperscript{3}) and coarse-grained (SONYC-L\textsuperscript{3}) resolutions show that the discriminative regions are concentrated in the lower frequency bands and accentuated in the coarser-grained representation}
	\label{fig_mels}
\end{figure}

\begin{table}[b]
	\centering
	\caption{SONYC-L\textsuperscript{3} uses a much coarser-grained input representation compared to L\textsuperscript{3}-Net or EdgeL\textsuperscript{3}}
	\vspace{-1mm}
	\resizebox{\columnwidth}{!}
	{\begin{tabular}{c|ccccc}
		\hline
		\textbf{Model} & \textbf{\begin{tabular}[c]{@{}c@{}}Sampling Freq.\\ (KHz)\end{tabular}} & \textbf{\begin{tabular}[c]{@{}c@{}}DFT\\ Size\end{tabular}} & \textbf{\begin{tabular}[c]{@{}c@{}}Num.\\ Mels\end{tabular}} & \textbf{\begin{tabular}[c]{@{}c@{}}Num.\\ Hops\end{tabular}} & \textbf{\begin{tabular}[c]{@{}c@{}}Num. Filters\\ (conv1/conv2)\end{tabular}} \\ \hline\hline
		L\textsuperscript{3}/EdgeL\textsuperscript{3}      & 48                                                                      & 2048                                                        & 256                                                          & 242                                                          & 64                                                                            \\
		SONYC-L\textsuperscript{3}       & 8                                                                       & 1024                                                        & 64                                                           & 51                                                           & 32                                                                            \\ \hline
	\end{tabular}}
\label{tab_input_reps}
\end{table}

\subsection{SONYC-L$^3$ Architecture}

The SONYC-L\textsuperscript{3} design is based on the observation that large embedding CNNs such as L\textsuperscript{3}-Net Audio are generally overparameterized for downstream tasks involving urban noise classification \cite{kumari2019edgel}, and therefore lend themselves effectively to compression strategies. 

\subsubsection{Coarser-grained input representation}

Through a comparison of full-resolution input melspectrograms of representative classes in SONYC-UST used by L\textsuperscript{3} versus their significantly coarser-grained counterparts used by SONYC-L\textsuperscript{3} (Table \ref{tab_input_reps}), we make two salient observations:

\begin{enumerate}[label=(\roman*),topsep=0pt,itemsep=-1ex,partopsep=1ex,parsep=1ex]
	\item Most of the energy is distributed in the relatively lower frequency bands for the majority of classes in UST. This is true even for alert signals (Figure \ref{fig_mels}(i)-(ii)) or human voice (Figure \ref{fig_mels}(iii)-(iv)): noise sources that are often perceived as being \textit{high-pitched}.

	\item Low spectral resolution, coupled with a reduced number of mel bins, accentuates the \textit{more discriminative} regions in the respective melspectrograms (Figure \ref{fig_mels}, column (ii)) when compared with their full-resolution counterparts.
\end{enumerate}

Thus, it might be possible to reduce the granularity of the melspectrograms significantly while still retaining enough information to discriminate the classes. This also has two added benefits: First, it reduces CNN's runtime activation memory. Second, since computing melspectrograms is a fairly heavy operation, a coarser representation is much more amenable to mote-scale realization.

\subsubsection{Reduced architecture}
In previous work, models such as EfficientNet \cite{tan2019efficientnet} have demonstrates the efficacy of uniformly scaling the network's width, depth, and image resolution to increase the network capacity. We follow a similar approach for downscaling L\textsuperscript{3}-Net Audio along the width dimension. 
\begin{figure}[t]
	\centering
	\includegraphics[width=1\linewidth, trim = 0 5 0 5, clip]{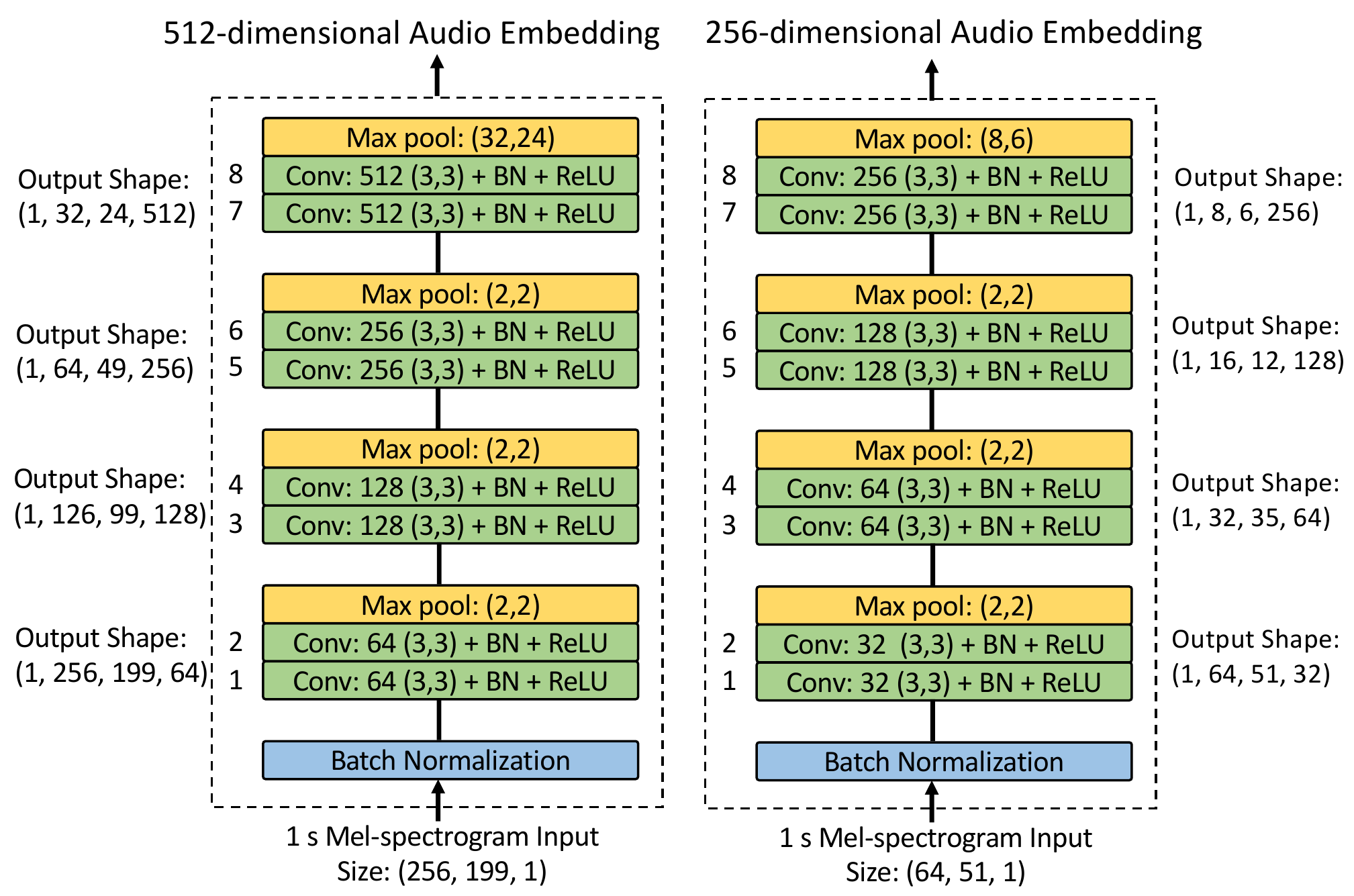}
	\caption{L\textsuperscript{3}-Net (left) and SONYC-L$^3$ (right) Audio Model}
	\label{fig:l3model}
	\vspace{-3.5mm}
\end{figure}

Specifically, we reduce the number of convolution filters in each \textit{conv} layer 
by 50\%, keeping the ratio of filters same as L\textsuperscript{3}. The final output is, therefore, a 256-dimensional embedding (Figure \ref{fig:l3model}).

\begin{table}[b]
	\centering
	\caption{SONYC-L\textsuperscript{3} activation memory is more than 100$\times$ lower than L\textsuperscript{3}-Net or EdgeL\textsuperscript{3}, and $\sim$30$\times$ lower than their quantized equivalents}
	\vspace{-1mm}
	\begin{tabular}{c|cc|c}
		\hline
		\multirow{2}{*}{\textbf{Layer}} & \multicolumn{2}{c|}{\textbf{L\textsuperscript{3}/EdgeL\textsuperscript{3}}}    & \textbf{SONYC-L\textsuperscript{3}}  \\ \cline{2-4}
		& \textbf{32-bit float} & \textbf{8-bit int} & \textbf{8-bit int} \\ \hline
		conv1/conv2                           & 12736                 & 3184               & 102                \\
		conv3/conv4                           & 6336                  & 1584               & 52                 \\
		conv5/conv6                           & 3136                  & 784                & 26                 \\
		conv7/conv8                           & 1536                  & 384                & 14                 \\
        \hline
\textbf{Total}                  & \textbf{47488}        & \textbf{11872}     & \textbf{388}       \\ \hline
	\end{tabular}
\label{tab_activ_mem}
\end{table}

When both these factors are combined with 8-bit quantization, the result is a model with activation memory that is more than 2 orders of magnitude lower than L\textsuperscript{3}-Net or EdgeL\textsuperscript{3} (Table \ref{tab_activ_mem}). And for static memory, it needs only 1.17 MB in 8-bit integer quantization, making it a feasible model for Cortex-M7.

\vspace{-1.5mm}
\subsection{SONYC-L\textsuperscript{3} Training: AVC vs. Specialized Embedding Approximation}
\label{sec_sonyc_l3}
As outlined in Section \ref{subsec_avc}, the same technique of using AVC to train L\textsuperscript{3}-Net can be applied to SONYC-L\textsuperscript{3}. However, the AVC setup doesn't allow us to leverage domain-specific data as the SONYC data does not collect multi-modal data. Additionally,  L\textsuperscript{3}-Net requires $\sim$11 TB of video data and up to 2 weeks on a 4-GPU HPC cluster to train both the audio and the video subnetworks, making it an expensive task to train for.



The above challenges can be addressed through \textit{Specialized Embedding Approximation} (SEA) \cite{srivastava2021specialized}, a knowledge distillation paradigm where the student (SONYC-L\textsuperscript{3}) is trained to only partially approximate the teacher's (L\textsuperscript{3}-Net) embedding manifold that is pertinent to the target domain of interest. More formally, given a teacher model $f_{\theta_T}(.)\in\mathbb{R}^n$, the technique aims to train a student $f_{\theta_S}(.)\in\mathbb{R}^d$ to mimic the teacher's embeddings on a new, unlabeled dataset $D_S$ representing the target domain (typically, $d<n$ for students intended to fit on mote-scale devices). Thus, the following objective function is optimized:
\begin{equation}
\min\limits_{\theta_S} \sum\limits_{x_j \in D_S} ||{f_{\theta_S}(x_j) - \phi(f_{\theta_T}(x_j))}||_2^2     
\end{equation}
where $\phi:\mathbb{R}^n\to\mathbb{R}^d$ is an appropriate dimensionality reduction function (such as PCA or UMAP \cite{mcinnes2018umap}) whose parameters are learned. 

The advantages of using SEA over AVC, or even traditional knowledge distillation, are two-fold. Firstly, SEA removes the reliance on the original training dataset by only requiring new, unlabeled data in the target domain; this is fairly easy to collect using deployed IoT sensors such as the SONYC MKI network. Secondly, it trains the student to learn only the portion of the teacher's manifold that is relevant to the target domain instead of its entire embedding space. Thus, it trades generality for training efficiency and achieves superior compression with far fewer data points. In fact, using SEA,  SONYC-L\textsuperscript{3} can be trained with \textit{an order of magnitude lesser data}, while also converging up to  $10\times$ faster than AVC training.


\subsection{SONYC-L\textsuperscript{3} Evaluation}
\label{sec_sonyc_l3_eval}
We evaluate SONYC-L\textsuperscript{3} on the SONYC-UST dataset using metrics described in Section \ref{sec_sonyc_ust} against the following baselines: VGGish \cite{jansen2017large}, L\textsuperscript{3}-Net \cite{arandjelovic2017look}, and EdgeL\textsuperscript{3} \cite{kumari2019edgel}.
For VGGish and L\textsuperscript{3}-Net featurizers, we train a multilayer perceptron (MLP) as well as a more sophisticated multiple-instance learning (MIL) classifier \cite{mcfee2018adaptive} with 0-2 hidden layers, and report the best results. Due to the lack of support for the MIL classifier's time-distributed convolutions in CMSIS-NN on Cortex-M7 devices and TensorFlow Lite on Raspberry Pis, both SONYC-L\textsuperscript{3} and EdgeL\textsuperscript{3} use only MLP classifiers.



\begin{table}[t]
	\centering
	\caption{Class-specific AUPRCs of SONYC-L\textsuperscript{3} vs baselines}
	\vspace{-3mm}
	\resizebox{\columnwidth}{!}
	{\begin{tabular}{l|cccHc}
		\hline
		\multicolumn{1}{c|}{\multirow{2}{*}{\begin{tabular}[c]{@{}c@{}} \textbf{Class}\\\textbf{ Label} \end{tabular}}} & \multicolumn{5}{c}{ \textbf{Model} }                                                                                                                                                                               \\
		\cline{2-6}
		\multicolumn{1}{c|}{}                                                                                           & \textbf{VGGish}  & \textbf{L\textsuperscript{3}-Net}  & \begin{tabular}[c]{@{}c@{}}\textbf{EdgeL\textsuperscript{3}}\end{tabular} & \begin{tabular}[c]{@{}c@{}}\textbf{SONYC-L\textsuperscript{3}}\\\textbf{-noquant}\end{tabular} & \textbf{SONYC-L\textsuperscript{3}}   \\
		\hline\hline
		engine                                                                                                          & 0.79              & 0.836            & 0.857                                                                     & 0.841                                                                     & 0.852               \\
		machinery-impact                                                                                                & 0.36              & 0.305            & 0.342                                                                     & 0.285                                                                     & 0.361               \\
		non-machinery-impact                                                                                            & 0.02              & 0.429            & 0.306                                                                     & 0.359                                                                     & 0.435               \\
		powered-saw                                                                                                     & 0.66              & 0.702            & 0.728                                                                     & 0.830                                                                     & 0.774               \\
		alert-signal                                                                                                    & 0.67              & 0.868            & 0.816                                                                     & 0.820                                                                     & 0.819               \\
		music                                                                                                           & 0.07              & 0.384            & 0.556                                                                     & 0.212                                                                     & 0.354               \\
		human-voice                                                                                                     & 0.84              & 0.959            & 0.945                                                                     & 0.951                                                                     & 0.950               \\
		dog                                                                                                             & 0.00              & 0.049            & 0.026                                                                     & 0.028                                                                     & 0.091               \\
		\hline
	\end{tabular}}
\label{tab_perclass_results}
\vspace{-1mm}
\end{table}

\subsubsection{Comparative performance on SONYC-UST}\hfill\\
\textbf{Per-Class Performance.}~~Table \ref{tab_perclass_results} lists the AUPRCs of each class in the multi-label SONYC-UST dataset for the compared baseline architectures. Out of these classes, the first five are potential sources of noise complaints and hence of interest to the SONYC project. To our surprise, we find that SONYC-L\textsuperscript{3} outperforms L\textsuperscript{3}-Net by 0.6$-$7.4\% on four out of these five classes. Interestingly, the AUPRC of a non-dominant class, \textit{dog}, is improved by almost 2$\times$. Thus, SONYC-L\textsuperscript{3} has a better multi-label classification performance even though it is constrained to use a weaker end classifier (MLP). The performance gains are even more significant when compared with VGGish, where up to 41\% improvement on per-class AUPRCs is observed. With respect to EdgeL\textsuperscript{3}, SONYC-L\textsuperscript{3} offers superior sensing for 6 out of the 8 classes, and improves the \textit{dog} class AUPRC by as much as 3.4$\times$.

\begin{table}[t]
	\caption{Micro-AUPRC, Micro-F1 and Macro-AUPRC of SONYC-L\textsuperscript{3} compared with the other baselines}
	\vspace{-3mm}
	\centering
	\resizebox{\columnwidth}{!}
	{\begin{tabular}{c|cccHc}
		\hline
		\multirow{2}{*}{ \textbf{Metric} } & \multicolumn{5}{c}{ \textbf{Model} }  \\
		\cline{2-6}
		& \textbf{VGGish}  & \textbf{L\textsuperscript{3}-Net}  & \begin{tabular}[c]{@{}c@{}}\textbf{EdgeL\textsuperscript{3}}\end{tabular} & \begin{tabular}[c]{@{}c@{}}\textbf{SONYC-L\textsuperscript{3}}\\\textbf{-noquant}\end{tabular} & \textbf{SONYC-L\textsuperscript{3}}   \\
		\hline\hline
		Micro-AUPRC                        & 0.77              & 0.810             & 0.791                                                                  & 0.781                                                                  & 0.785               \\
		Micro-F1 (0.5)                     & 0.70              & 0.723            & 0.716                                                                  & 0.712                                                                  & 0.702               \\
		Macro-AUPRC                        & 0.43              & 0.566            & 0.572                                                                  & 0.541                                                                  & 0.579               \\
	\hline
	\end{tabular}}
	\label{tab_overall_results}
	\vspace{-5mm}
\end{table}

\vspace{1mm}
\noindent
\textbf{Overall Performance.}~~Table \ref{tab_overall_results} outlines the performance of SONYC-L\textsuperscript{3} on the three metrics of interest in the SONYC-UST dataset. While it outperforms all compared baselines on macro-AUPRC, the difference in the other metrics is $<$2.5\%. This, coupled with the fact that SONYC-L\textsuperscript{3} is 3$\times$ more efficient than either baseline from a runtime perspective, $\sim$2 orders of magnitude smaller than L\textsuperscript{3}-Net and uses $1.2$ orders of magnitude less active memory than either baseline, it is clear that our proposed solution is the most adept at navigating the accuracy-efficiency tradeoff.

The per-class analysis also explains the apparent anomaly in Table \ref{tab_overall_results}, where SONYC-L\textsuperscript{3} improves upon L\textsuperscript{3} in terms of macro-AUPRC but underperforms slightly on micro-AUPRC and F1 (all three metrics are improved over VGGish by up to 15\%). The micro-averaged metrics are weighed by the distribution of the class labels, and hence are biased towards the over-represented classes in the unbalanced SONYC-UST dataset. In particular, the underperformance of SONYC-L\textsuperscript{3} on the \textit{alert signal} class skews the micro-averaged metrics in favor of the baseline, even though SONYC-L\textsuperscript{3} is the overall superior solution. Analogous reasoning can be formulated for both variants of EdgeL\textsuperscript{3} with regards to the \textit{engine} class, which has the highest predominance in the UST test set.

\section{Infrastructure-free, Low-cost WSN}\label{sec:inf_agg}

The cost of mounting infrastructure connected devices on poles or intersections is on the order of \$1000-2000 per device. 
Additionally, the process of getting permissions and qualified labor to mount such devices is onerous and delay prone. Anecdotally, these issues have materially impacted smart city wireless sensor deployments. MKII mitigates these issues by being deployable without needing wall power or existing network infrastructure.

Further, we claim that the MKII has sufficient functionality to compete on a cost basis with more expensive commercial noise and SPL monitoring equipment (and their service contracts) with reasonable data quality. Existing high-end longitudinal noise monitoring solutions typically cost in excess of \$10K/sensor with considerable annual upkeep fees that can run into the millions with larger deployments, reducing their scalability and viability, except in certain state level initiatives such as aircraft noise monitoring \cite{webtrak_web}. The MKII system has the capability to provide a comparable service, while also delivering enhanced features such as automatic source ID at a significantly lower cost over time.

\vspace*{-3mm}
\subsection{Self-Powered MKII Device Design}
While several systems details of the device design, including its supports for compute performance, SPL meter, and timing are relegated to the Appendix, we focus here on the low power aspects of the design that support truly wireless operation. 

\begin{figure}[t]
   
	\centering
		\begin{subfigure}[h]{0.188\textwidth}
		\centering
		\includegraphics[width=\textwidth]{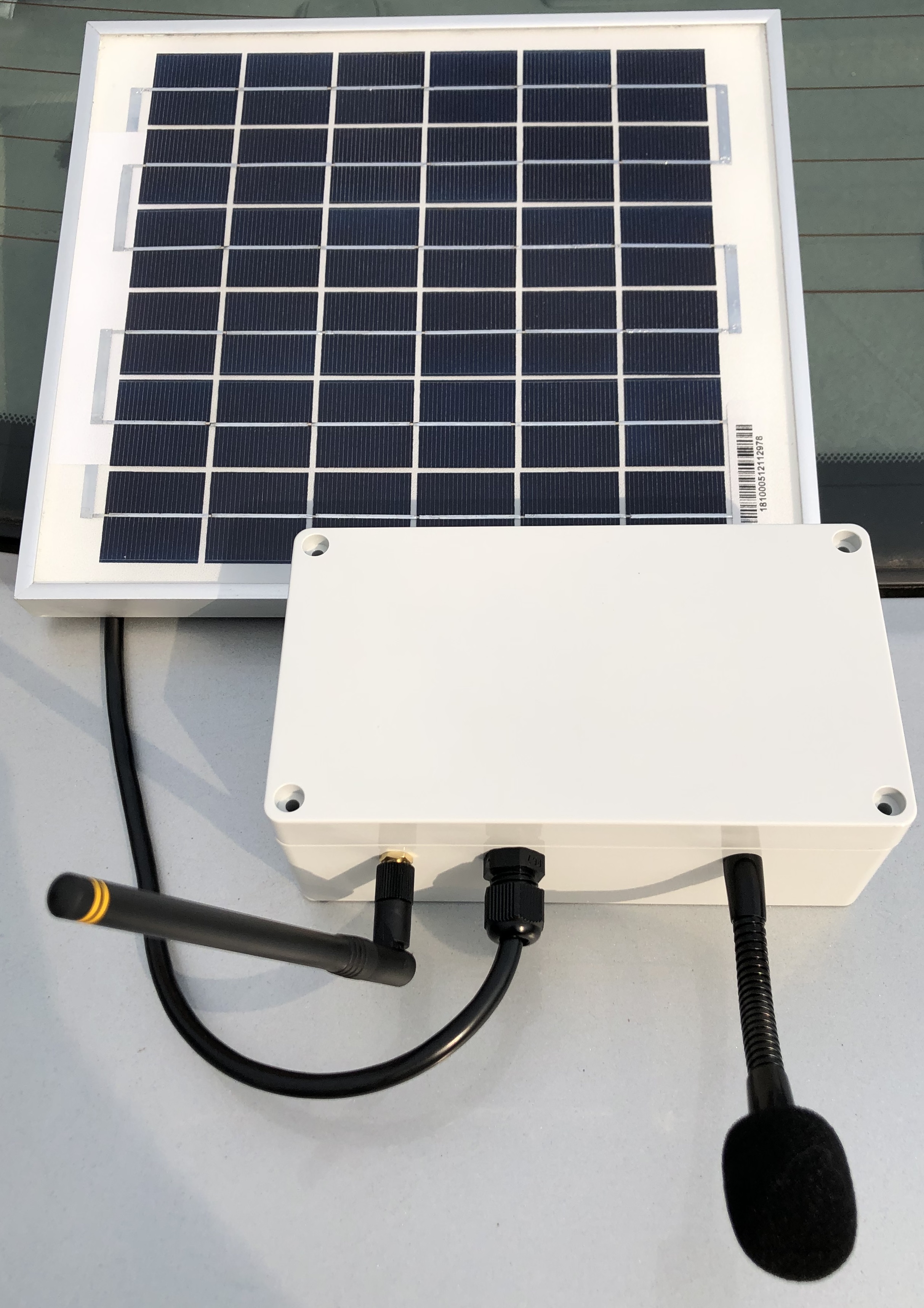}
			\caption{Assembled Unit}
		\end{subfigure}%
		\begin{subfigure}[h]{0.211\textwidth}
			\centering
            \includegraphics[width=\textwidth]{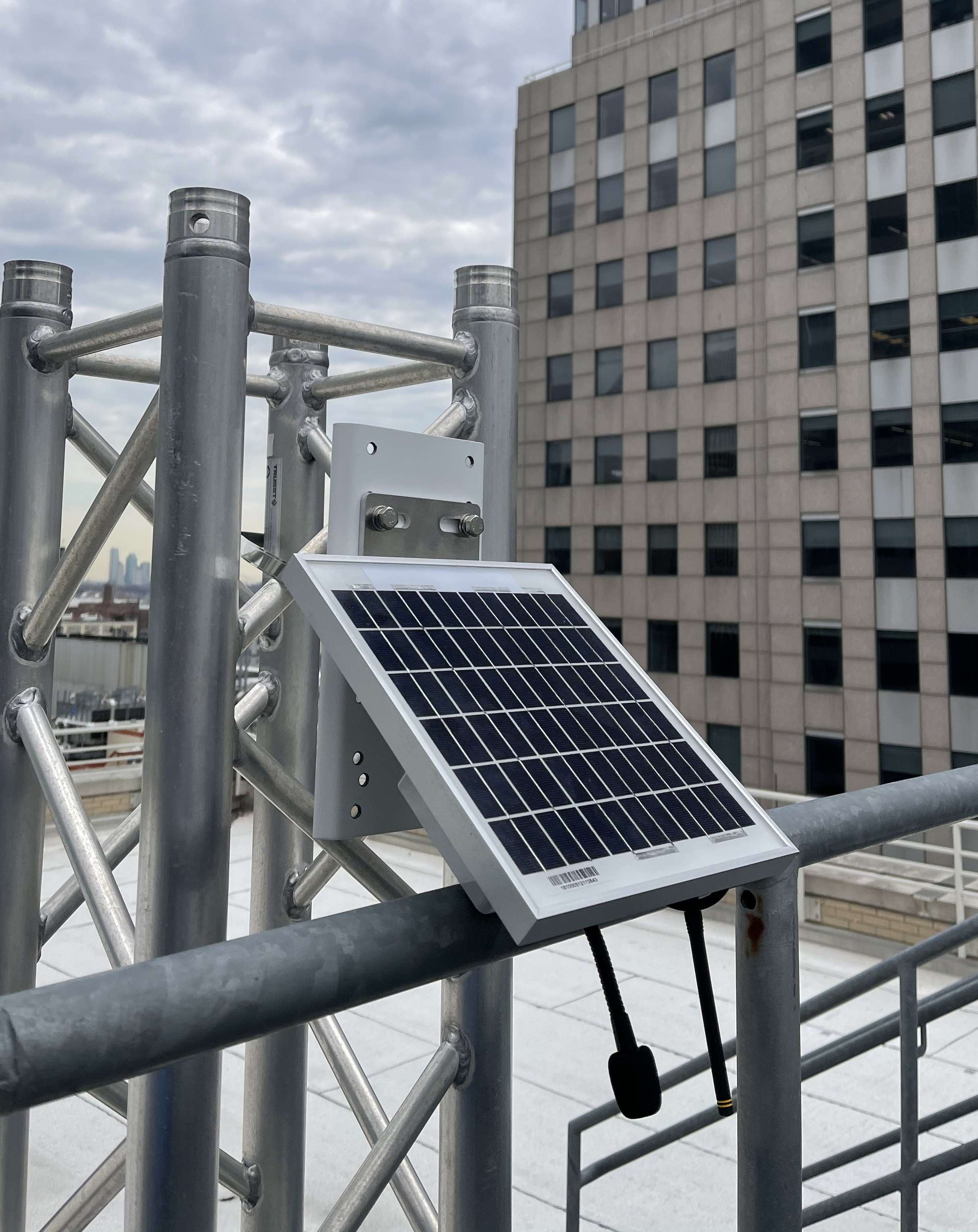}
			\caption{Deployed in Tier 2}
		\end{subfigure}%
~		 \vspace{-2mm}
	\caption{The SONYC MKII device, shown in enclosed form, is affixed behind the solar panel, which is angularly mounted to maximize solar exposure}
	\label{fig:mk2}
	\vspace{-2mm}
\end{figure}

\vspace{1mm}
\noindent
{\bf Solar Energy Harvester}.~~~At the outset of the design we targeted a power budget of 100 mW for applications running on the MKII hardware and a goal of no more than one power outage per year due to below average solar inputs. From analyzing historical data from the National Centers for Environmental Information (NCEI) we concluded that buffering at least three days worth of energy would meet this requirement. These considerations informed the power capacity of the solar harvester to deal with variable solar input and provided a basis for selection of a small (8.75"x10.63") 5 W solar harvester (Figure \ref{fig:mk2}) and lithium-titanate batteries to meet the requirements safely (as elaborated in the Appendix).  

\vspace{1mm}
\noindent
{\bf Power Management}.~~~The complex power management needs for MKII---including flexibility in battery chemistry and needing to manage solar harvesting as well as batteries---precluded the use of a typical integrated battery management chip. Instead, a fully software-defined battery management system (BMS) was programmed to a separate Cortex-M3 processor based MCU. A small independent system was desirable because a power manager that was itself low-power was needed to avoid loading the system in scenarios of excess depletion. Further, the BMS is computationally simple but I/O (pin count) heavy while a pin on the STM32H753 is conceptually expensive. It was also desirable from a systems perspective because the BMS services are critical and always-on, warranting a degree of isolation. 
With an independent BMS the STM32H753 can be safely wiped, debugged, or crash without risk of damaging the battery or otherwise interrupting vital services. The core BMS tasks consist of battery management (including balancing via selective discharging and overcharge avoidance), solar harvester management, 
on-the-fly reconfiguration to use USB power input when plugged in, and reporting monitored power data to the H7. 

\vspace{1mm}
\noindent
{\bf Power Analysis}.~~~Each inference invocation (consuming 1-second of audio data) requires about 750 ms of execution time at the maximum STM32H753 clock frequency of 480 MHz, which all but dominates the power budget. The microphone, acquisition, and front-end processing consumption is about 6~mW, and the networking consumption is typically less than 15~mW. Operations outside of inference are run at reduced clock speed to save power. All told, our targeted total power budget of 100~mW is thus approximately met (i.e, at 107~mW) by running the inference at 25\% duty cycle, which is acceptable for the SONYC application.
\subsection{Instantiating and Validating LoRa Links}
LoRa is a wireless physical layer protocol based on Chirp Spread Spectrum (CSS), which is reported to have 21 km range in Line of Sight (LoS) environments and up to 2 km range in Non Line of Sight (NLoS) urban environments though several (4-6) building with +14 dBm output power \cite{libelium}.  While using the same LoRa chip as that report, our design supports a total output power of 27 dBm, which is within the FCC mandated limit of 30 dBm.  We operate in the 902-928 MHz ISM band.  

Prior to finalizing the MKII design, we did an empirical study to corroborate the achievable LoRa range in urban settings. We conducted over two dozen LoRa link tests to evaluate the performance of LoRa in both LOS and NLoS conditions, with different types of clutter (parks, downtown areas, suburban areas) in Manhattan and Brooklyn in NYC and elsewhere during different weather/seasonal conditions. Metrics of Packet Reception Rate (PRR) and Receive Signal Strength Indication (RSSI) were computed for various LoRa configurations at distances ranging from 300m to just over 2000m, between LoRa TX and RX node pairs typically placed atop a 10 ft pole.  In each test, 200 packets were transmitted per frequency, for a total of 10 (and in a few cases 52) frequencies. 

Tables \ref{tab_LoS_test_results} and \ref{tab_NLoS_test_results} respectively show the performance in Line-of-Sight and non Line-of-Sight environments, with medium power (19 dBm) and high power (27 dBm) links. Based on these tests and our choice to configure the radio with a bandwidth of 500KHz, spreading factor of 8, coding factor of 4$/$5, and power of 27 dBm, we conservatively estimate 500m links to be reliably realized in the varied city clutters of SONYC deployment, albeit in most cases we expect the achievable link length to be substantially higher. 
  
\begin{table}
  \caption{Packet Reliability and RSSI in LoS Tests} 
  \label{tab_LoS_test_results}
   \vspace{-3mm}
  \centering
  \begin{adjustbox}{max width=\columnwidth}
  \begin{threeparttable}
    \begin{tabular}{l| c c c c | c c}
      \hline
		\textbf{Location} & 
		\textbf{\begin{tabular}[c]{@{}c@{}}LoRa\\Configuration\\(BW, SF, CR)\end{tabular}}& 
		\textbf{\begin{tabular}[c]{@{}c@{}}Power\\ Level\\(dBm)\end{tabular}}& 
		\textbf{\begin{tabular}[c]{@{}c@{}}Link\\ Range\\(m)\end{tabular}}&
		\textbf{\begin{tabular}[c]{@{}c@{}}Data\\ Rate\\(kbps)\end{tabular}}&
		\textbf{\begin{tabular}[c]{@{}c@{}}Average\\ PRR\end{tabular}} & 
		\textbf{\begin{tabular}[c]{@{}c@{}}Average\\ RSSI\\(dBm)\end{tabular}} \\
    	\hline
    	\hline
		  1: City Rooftop & (250, 8, 4/6) & 19 & 2160 & 5.208 & 0.817 & -117.02\\ 
		  2: Downtown Park & (250, 8, 4/6) & 19 & 1000 & 5.208 & 0.985 & -109.00\\ 
		  2: Downtown Park & (500, 7, 4/5) & 19 & 1000 & 21.875 & 0.968 & -108.52\\
		  3: Corn Field & (250, 8, 4/6) & 19 & 500 & 5.208 & 0.998 &  -101.19\\ 
		  3: Corn Field & (500, 7, 4/5) & 19 & 500 & 21.875 & 0.998 &  -99.94\\
		  4: Sports Grounds & (500, 7, 4/5) & 19 & 500 & 21.875 & 0.983 &  -110.33\\ 
		  5: Parking Lot & (500, 8, 4/5) & 27 & 400 & 12.5 & 0.995 & -82.24 \\ 
		 \hline
     \end{tabular}
  \begin{tablenotes}
  \end{tablenotes}
  \end{threeparttable}
  \end{adjustbox}
  \vspace{-1mm}
  \end{table}

\begin{table}
  \caption{Packet Reliability and RSSI in NLoS Tests} 
   \vspace{-3mm}
  \label{tab_NLoS_test_results}
  \centering
  \begin{adjustbox}{max width=\columnwidth}
  \begin{threeparttable}
    \begin{tabular}{l| c c c c | c c}
      \toprule
		\textbf{Location} & 
		\textbf{\begin{tabular}[c]{@{}c@{}}LoRa\\Configuration\\(BW, SF, CR)\end{tabular}}& 
		\textbf{\begin{tabular}[c]{@{}c@{}}Power\\ Level\\(dBm)\end{tabular}}& 
		\textbf{\begin{tabular}[c]{@{}c@{}}Link\\ Range\\(m)\end{tabular}}&
		\textbf{\begin{tabular}[c]{@{}c@{}}Data\\ Rate\\(kbps)\end{tabular}}&
		\textbf{\begin{tabular}[c]{@{}c@{}}Average\\ PRR\end{tabular}} & 
		\textbf{\begin{tabular}[c]{@{}c@{}}Average\\RSSI\\(dBm)\end{tabular}} \\
		\hline
		\hline
		  7: Campus Oval & (250, 8, 4/6) & 19 & 500 & 5.208 & 0.956 & - 109.78 \\ 
		  7: Campus Oval & (500, 7, 4/5) & 19 & 500 & 21.875 & 0.922 & -114.99 \\
		  8: Campus Oval & (500, 8, 4/5) & 27 & 493 & 12.5 & 0.987 &  -90.06 \\
		  8: Campus Oval & (500, 8, 4/5) & 27 & 493 & 12.5 & 0.981 &  -87.68 \\
		  9: Urban Street & (250, 8, 4/6) & 19 & 500 & 5.208 & 0.764 & -120.05 \\ 
		  10: Downtown Street & (250, 8, 4/6) & 19 & 300 & 5.208 & 0.917 & -110.28 \\ 
		 \bottomrule
     \end{tabular}
  \begin{tablenotes}
  \end{tablenotes}
  \end{threeparttable}
  \end{adjustbox}
    \vspace{-3mm}
  \end{table}

\vspace{.1mm}
\noindent
\subsubsection{FCC Compliance.}
While our networking desiderata of achievable range, data rate, and power consumption are the primary guide for the selection of LoRa radio supported configurations of its physical layer (in terms of center frequency, spreading factor, coding rate, and bandwidth), FCC requirements \cite{FCC15.247} are also a key factor.

To comply with FCC 15.247 \cite{FCC15.247}, our medium access protocol design has to choose between a single frequency and several multi frequency access modes. The latter modes require selecting 25 or more frequencies per node and impose dwell time upper bounds per frequency, which lead to significantly higher overhead.  The former requires a frequency width of at least 500 KHz, but does not prevent nodes from changing their chosen frequency over time nor from using different frequencies from each other. We chose the former, with a physical layer bandwidth configuration of 500 KHz; to balance between range and power consumption, we selected spreading factor to 8, which yields a receiver sensitivity of -121 dBm; and to constrain power consumption and transmission length, we chose a modest coding rate of 4$/$5.  The net result of this LoRa configuration is a link rate of 12.5kbps.

\subsection{Multi-hop, Low-power LoRa Network}
Meeting the coverage requirement of up to 5km end-to-end range with reliable >500m links led us to realize a multi-hop Tier 2 network, as opposed to the star-of-stars topology supported by the  LoRaWAN standard. Our network design incorporates atop the LoRa PHY layer a MAC protocol, a convergecast routing protocol for sensing inferences and data from MKII nodes to their gateway MKII node, and a sort of flooding protocol in the reverse direction (for configuration, command and control).  It has been tested to reliably handle the expected traffic across a significant number of hops ($\sim$7) in this and previous projects.  In the common case, though, we expect that SONYC ad hoc Tier 2 deployments will be 2-3 hops.


Even though transmission consumes over 2W, our network design only consumes 15mW power consumption overall. This is achieved via a duty cycled MAC protocol. Given the limited rate of both data and control messages, the MAC optimizes for the receiver power by aggressive (1.1\%) duty-cycling using a synchronous, receiver-centric protocol, OMAC \cite{omac, configurationspace}, where each receiver shares its respective pseudo random wakeup times with with its neighbors, which are asynchronously discovered, via a pseudo-random seed.  With an appropriate choice of duty cycle, OMAC eschews self-interference within the network.

\section{Dynamic Management of External Interference And Link Fading}

In the cluttered wireless environment of the 902-928 MHz ISM band in NYC, management of external interference and link fading turns out to be the critical networking challenge that we need to address. To characterize external interference and link fading, we conducted data collection campaigns at multiple locations in NYC and elsewhere prior to deployment, as well as in the Tier 2 network that we deployed, each ranging from hours to a week. Data was collected that measured frequency noise passively, both with the MKII devices and a LimeSDR \cite{limesdr} device, as well as PRR, SNR, and RSSI actively for links between MKII node pairs.

\begin{figure}[t]
	\centering
	\includegraphics[width=0.42\textwidth, clip]{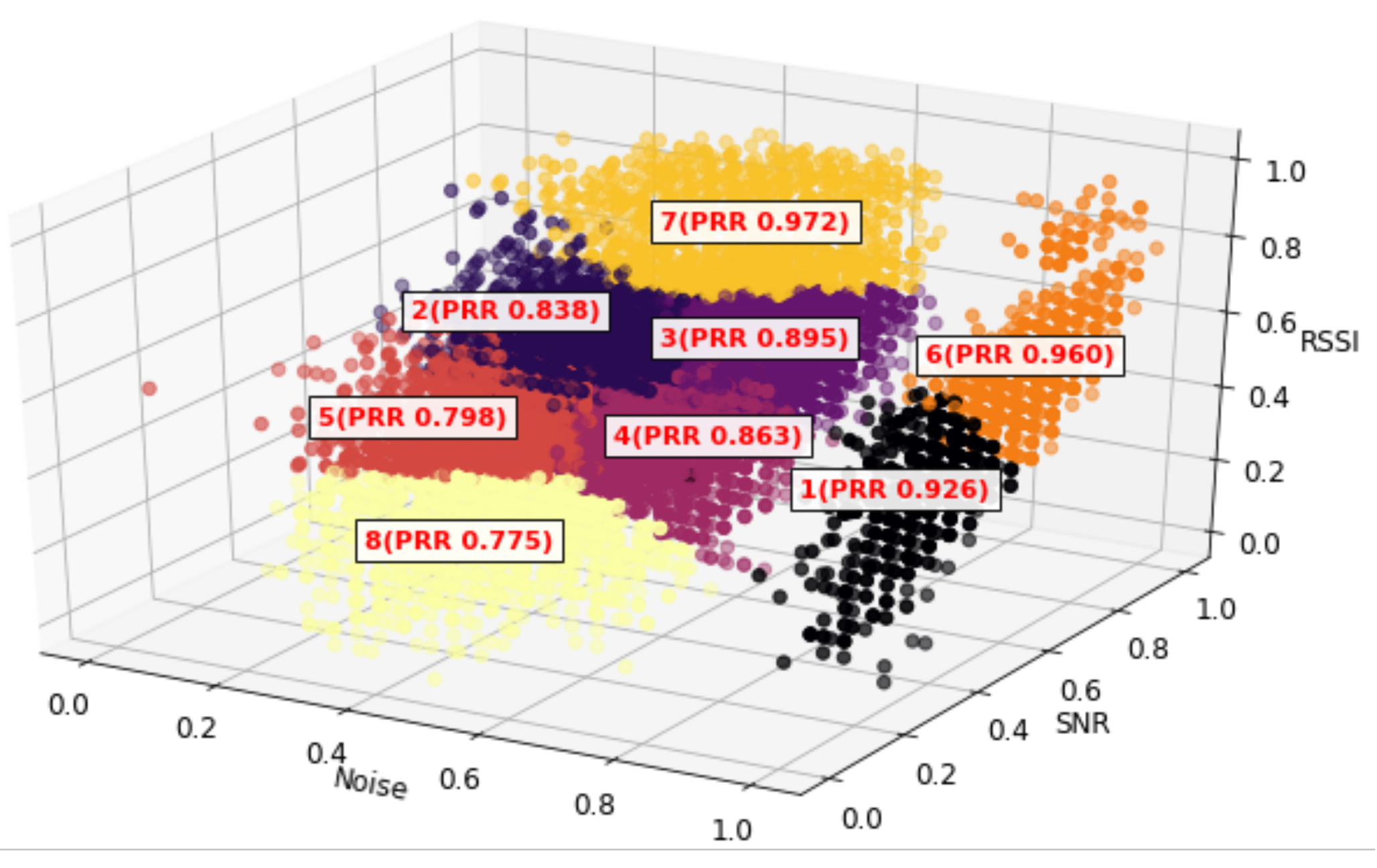}
	\vspace{-3mm}
	\caption{A clustering of min-max normalized noise, SNR, RSSI metrics over link  time intervals collected from Tier 2 network. Clusters 1 and 6 indicate the existence of fading link time intervals}
	\label{cluster_fading}
\vspace{-3mm}	
\end{figure}

Figure \ref{cluster_fading} illustrates the existence of external interference as well as fading in links. It depicts a clustering of link metrics (over noise, SNR, RSSI, and PRR measures) taken from our Tier 2 network.  Each point corresponds to metric data collected from some link over a 10 s interval.  
With respect to external interference, cluster 8 stands out: it has rather high noise (typically around -9x dBm) and low SNR values, which indicate the presence of external interference and explains the low PRR.  (An intuitive visualization of external interference may be seen in the Appendix.)  With respect to fading, clusters 1 and 6 stand out: they have rather low noise (typically around -125dBm) compared to the other clusters that indicates a low interference regime. Even when the SNR and RSSI of the received packets is quite good in these points, packets are lost with significant probability, pointing to the existence of fading during these link time intervals.

Fortunately, the empirical data from MKII networks in different geographic locations shows that in each network their exist frequencies that yield high packet reliability, both over time and over multiple links. More generally, in these and other data (cf.~the Appendix) we find that {\em while the frequency with the best PRR need not be the same for all links across a large geographic region, a ``good'' common frequency often exists across limited geographic regions}.


\subsection{Offline and Online Management}

We now present three methods for selecting a common frequency that avoids external interference and link fading. One of these methods is performed offline and then used to configure the frequency it chooses in a Tier 2 network under deployment. This Offline method involves collecting short-term time interval metrics from links that are established in pre-deployment data collection campaigns at intended or surrogate locations.  From the metrics in the collected data, the average of long-term PRR over the surrogate links is used to rank and accordingly select the top frequency for configuration.

 \begin{figure}[t]
    \vspace{1mm}
 	\centering
	\includegraphics[width=0.47\textwidth]{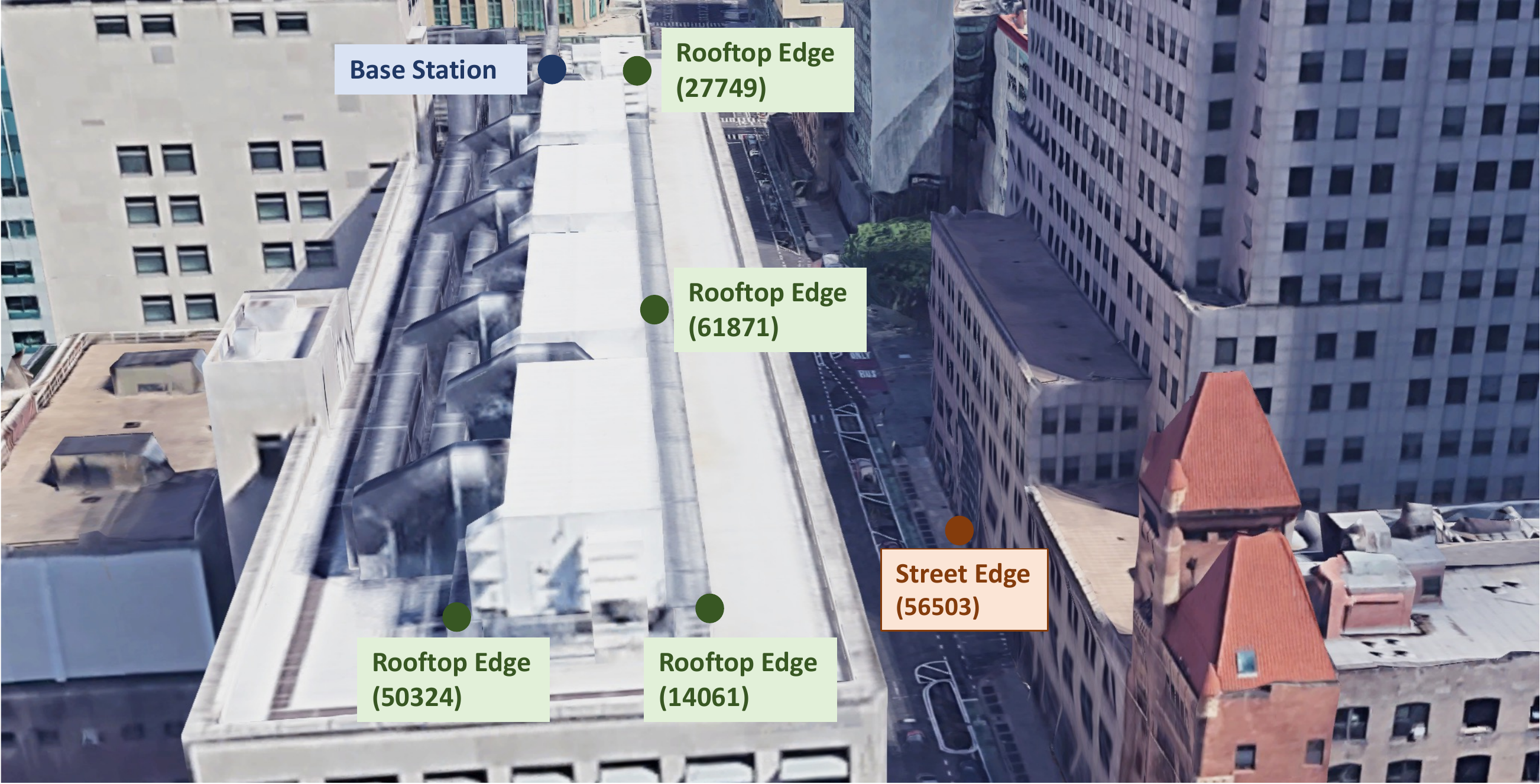}
	 \vspace{-2mm}
	\caption{Tier 2 network deployed in a rooftop and street-level setting in downtown Brooklyn} 
	\label{fig:deployment}
\end{figure}

The other two methods are performed online for in situ frequency selection, on a per network or per node basis. In the first Online method, (say 95\%ile) noise, (5\%ile) SNR, (5\%ile) RSSI, and interval PRR, is computed for a specified number of short-term intervals for each frequency.  Over all links under consideration ---which may be network-wide or incident at any node--- the sum of the per-link per-time interval average of the sum of min-max normalized noise, SNR, RSSI, and short-term PRR is computed to rank and choose the top frequency for the upcoming period.

This online method is communication intensive, so for our low power solution we approximate it via a Low-power Online version, which sequentially performs passive metric estimation (on noise) first for all frequencies, downselects the top-k frequencies with the best min-max normalized (say 95\%ile) noise and then, similar to its parent method, computes the  the sum of the per-link per-time interval average of the sum of min-max normalized noise, SNR, RSSI, and short-term PRR to rank and choose the top frequency for the upcoming period.



Our online methods allow frequency adaptation to be provided on an ongoing basis and to be performed either locally or globally:  In the local case, each node locally chooses the best frequency across of its incoming links as its receive frequency.  OMAC by virtue of being a self-stabilizing receiver-centric protocol implicitly coordinates the use of this frequency with its neighboring nodes, with little modification to its protocol. In the global case, the network manager convergecasts the active measures from all nodes to the Tier 1 base station, which chooses the frequency with the best average of the measures across all links, and wirelessly reconfigures the Tier 2 network in situ. We support both cases, since the local adaptation is likely to fine tune the chosen frequency to improve reliability more effectively on a per node basis.  Conversely, the global adaptation has the advantage that the selection of the network-wide frequency can be combined with any pre-knowledge of frequencies to avoid and/or programmatic coordination with external networks, which is a recommended practice. (Note that we eschew adaptation of the frequency that is used for discovery: this is to achieve fault-tolerance, so that nodes that are inadvertently not up during adaption are not excluded from the WSN.)

\section{Brooklyn Deployment Results}

 Figure \ref{fig:deployment} depicts our Tier 2 WSN deployment in downtown Brooklyn, with MKII nodes on the roof of a tall building as well as at street level.  Following an extended testing period, we have collected and evaluated over the last three months acoustic and network data from this WSN, in various configurations and numbers/locations of MKIIs.  Grossly, the Tier 2 network has remained up over this period, albeit there have been a few periods of data loss in the Tier 1 node interfaced to the WSN, because of emergent stability issues in that node's wired infrastructure network connection and operator errors in configuring its data collector tools.

\begin{table}[!tb]
	\caption{Top-5 frequencies selected by different methods: offline, exhaustive online, and low-power online, with their corresponding metric values}
	 \vspace{-3mm}
	\centering
    \begin{adjustbox}{max width=\columnwidth}
	\begin{tabular}{c|cccccc}
		\hline
        \textbf{Method} &
		\multicolumn{6}{c}{\textbf{Top 5 Frequencies (MHz)}} \\
	    \hline\hline

	    {\multirow{2}{*}{\textbf{\begin{tabular}[c]{@{}c@{}}Offline\\ \end{tabular}}}}
	    & 918.5 & 908 & 922.5 & 912 & 910.5 & All others \\ 
	    \cline{2-7} &  0.9976 & 0.9973 & 0.9972 & 0.9971 & 0.9970 & 0.8961\\
	    \hline
	    
	    {\multirow{2}{*}{\textbf{\begin{tabular}[c]{@{}c@{}}Online \end{tabular}}}}
	    & 918.5 & 910.5 &  917.5 &  903 & 912 & All others\\ 
	    \cline{2-7} & 0.8277 &	0.8243 & 0.8167 &	0.8129 &	0.8124 & 0.7381\\
	    \hline
	  
	    {\multirow{2}{*}{\textbf{\begin{tabular}[c]{@{}c@{}}Low-power\\Online \end{tabular}}}}
	    & 918.5 & 910.5 & 917.5 & 910 & 918 & All others \\
	    \cline{2-7} & 0.8277 &	0.8243 & 0.8167 & 0.8102 &	0.8094 & 0.7394\\
	    \hline
	    
	\end{tabular}
  \label{tab_freq_select}
  \end{adjustbox}
\vspace{-2mm}
\end{table}

\begin{table}[b]
	\caption{Validation of frequency selection methods}
	\vspace{-3mm}
    \centering
   \begin{adjustbox}{max width=\columnwidth}
	\begin{tabular}{c|c|cc|c|c|c}
		\hline
		{\multirow{2}{*}{\textbf{\begin{tabular}[c]{@{}c@{}}Frequency\\ (MHz) \end{tabular}}}} &
	    {\multirow{2}{*}{\textbf{\begin{tabular}[c]{@{}c@{}}Ground-\\truth PRR \end{tabular}}}} &
		{\multirow{2}{*}{\textbf{\begin{tabular}[c]{@{}c@{}}Test1 \\PRR \end{tabular}}}} &
		{\multirow{2}{*}{\textbf{\begin{tabular}[c]{@{}c@{}}Test2 \\PRR \end{tabular}}}} &
		\multicolumn{3}{c}{\textbf{Top 5 Selection}} \\
		\cline{5-7} & & & &
		\textbf{\begin{tabular}[c]{@{}c@{}}Offline \end{tabular}} & \textbf{\begin{tabular}[c]{@{}c@{}}Online \end{tabular}} & \textbf{\begin{tabular}[c]{@{}c@{}}Low-power\\Online \end{tabular}}
		\\ \hline\hline
		918.5 & 0.9976 & 0.9951 & 0.9951 & 1 & 1  & 1 \\
		908  & 0.9973 & 0.9946 & 0.9965 & 2 & & \\
		922.5 & 0.9972 & 0.9941 & 0.9926 & 3 & & \\
		912  & 0.9971 & 0.9946 & 0.9962 & 4 & 5 & \\
		910.5  & 0.9970 & 0.9960 & 0.9959 & 5 & 2 & 2 \\
		903 & 0.9950 & 0.9951 & 0.9934& & 4 & \\
		917.5 & 0.9863 & 0.9942 & 0.9174 & & 3 & 3\\
		910  & 0.9957 & 0.9939 & 0.9947 & &  & 4\\
		918 & 0.9967 & 0.9941 & 0.9938 & &  & 5\\
        \hline
        All others & 0.8866 & 0.8991 & 0.8924 & & & \\
        \hline
	\end{tabular}
\label{tab_freq_verify}
\end{adjustbox}
\end{table}
 
 We  validated the performance of three frequency selection methods, to show that these methods are effective in (near) optimization of selection of the best frequency for their (training) period of data collection but also for upcoming (test) period. Tables \ref{tab_freq_select} and \ref{tab_freq_verify} show that all three methods select the same top frequency, 918.5 MHz, from the 52 available frequencies, which has the highest ground truth PRR (of 99.76\%) over the method training period and which continues to have very high PRR in a couple of subsequent method testing periods. The Online methods, while not producing the identical (and the ideal) top-5 frequencies chosen by the Offline methods, still agree on several other frequencies in their top-5 frequencies, and even the remaining frequencies have relatively high PRR values. We note that the third-best selection of both Online methods is less than ideal; this is attributed to the limitation of our chosen ranking metric for capturing link fading with its short-term metrics, suggesting that long-term PRR should be used instead of short-term PRR.
 
One lesson from this deployment has been the importance of per-node duty cycle selection relative to its neighborhood degree. Our de facto configuration of a 1.1\% receiver duty cycle is optimized to avoid self-interference in the network when nodes have 2-3 neighbors; this node degree is both sufficient to tolerate some node loss while maintaining network connectivity as well as to yield affordable coverage.  But this configuration predictably underperforms as the degree increases, as shown in Table \ref{rooftop_network}, since the traffic sourced by each node is relatively regular and so self-interference becomes more likely as the degree increases.  This motivates the need for a formulaic increase in the duty cycle, incurring more power consumption for networking, although not in a way that significantly affects the overall system.



\begin{table}[t]
    \small
	\centering
	\caption{Impact of density increase on reliability if node duty cycle remains unchanged}
	\vspace{-3mm}
	\begin{tabular}{c|c|c|c}
		\hline
		\textbf{\begin{tabular}[c]{@{}c@{}} Edge Degrees\\ \end{tabular}} &
		\textbf{\begin{tabular}[c]{@{}c@{}} Total Sent \\ \end{tabular}}&
		\textbf{\begin{tabular}[c]{@{}c@{}} Total Received \\\end{tabular}} &
		\textbf{\begin{tabular}[c]{@{}c@{}} PRR \end{tabular}} \\ \hline\hline
		2  & 102217 & 100142 & 0.9797 \\
		\hline
        4 &  62421 & 55084 & 0.8825 \\
        \hline
        5 & 42601 &	29085 & 0.6915 \\
        \hline
	\end{tabular}
\label{rooftop_network}
\vspace{-2.5mm}
\end{table}

\begin{figure}[tbh]
 	\centering
	\includegraphics[width=0.47\textwidth]{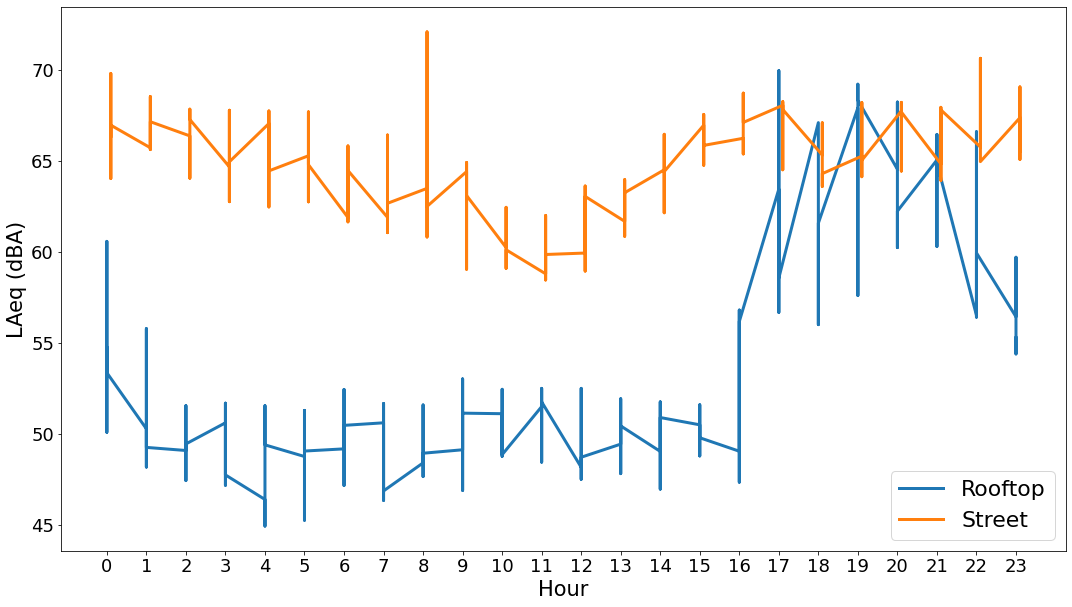}
	\vspace{-2mm}
	\caption{Average sound level (LAeq) for edge nodes at the rooftop and street level deployment}
	\label{fig:laeq}
\end{figure}
\begin{figure}[tbh]
 	\centering
	\includegraphics[width=0.47\textwidth]{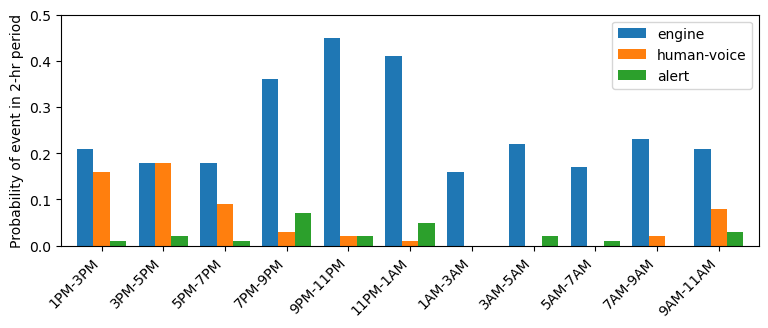}
	\vspace{-4mm}
	\caption{Likelihood of occurrence of \textit{engine}, \textit{human-voice} or \textit{alert} noise source in contiguous 2-hour periods} 
	\label{fig:day_analysis}
\end{figure}

Figures \ref{fig:laeq} and \ref{fig:day_analysis} respectively illustrate the SPL and classification features of the application based on data collected. The former showcases the difference in noise level at different heights: \textit{rooftop} (20ft) and \textit{street} (ground) level. We observe that the downtown street level, which is also close to the railway line, is consistently being somewhat loud.
We leverage the street level deployment to further analyze the evolution of 3 sound event classes: \textit{engine}, \textit{human-voice} and \textit{alert} over a continuous period of time. Fig. \ref{fig:day_analysis} depicts the probability of occurrence of a noise source in contiguous 2-hour periods. Due to its location at the street level and proximity to the rail line, \textit{engine} class is the most likely event to occur with an average probability of occurrence of $0.25$. Because the data corresponds to a weekday (Wednesday), it's not surprising that \textit{human-voice} drops after 5PM and then rises again after 7AM. In contrast, \textit{alert} class, which is dominant by the presence of sirens, does not have any deterministic pattern.

\section{Discussion}

\vspace{1mm}
\noindent
\textbf{In situ programmability}.~~~Like the SONYC MKI system, MKII also supports a rich management and reconfiguration module for configuring various system parameters locally as well as remotely, such as the ML duty cycle, SPL threshold to run the CNN, decision aggregation and status update periods, calibration compensation filter configurations, frequency selection, etc. While eMote does support reprogramming of managed apps in the field, this capability has not been integrated in the MKII version awaiting fielding. The case for updating native software components such as the CNN model in the field remains a relevant consideration for future work, as we expect active learning and federated learning support at the mote level to become increasingly important for long lived systems.  This suggests that we either refine the eMote interpreter via a JIT compiler to reduce the slowdown incurred in the VM (especially for compute-intensive tasks) and migrate these components to the managed level, which has implications for ML synthesis and instrumenting toolchain discussed earlier.


\vspace{1mm}
\noindent
\textbf{Limited network capacity}.~~~While the LoRa network offers good coverage and low power consumption, its achievable data rates in LoRa can be limiting for smart city applications that require continuous streaming, frequent data collection or large model updates.  Those apps require substantially higher capacity networks which typically require more infrastructure support and have very different coverage considerations.

We chose to operate links at 12.5kbps, for reasons discussed earlier. 
As instrumented, the effective MKII edge-to-gateway rate is only a small fraction of that, since OMAC's duty cycling emphasizes energy efficiency and not throughput maximization per se.  In a modest size network of 20 nodes with up to 3 hops, each node can be expected to source at most {\small $\sim$}100KB per day.  For achieving higher rates, OMAC can be refined, by rate adapting its links and by changing its globally fixed duty cycle to an adaptive duty cycling per node, while avoiding network self-interference. This could yield about several times higher traffic. But, even with 1MB capacity per node per day, MKII networks refined thus would be relevant primarily for applications which embrace in situ computing in lieu of communication.  

\section{Related Work}

\vspace{1mm}
\noindent
\textbf{Mote-scale deep learning}.~~~To maximize the performance and minimize the memory footprint of neural networks on Cortex-M processors, ARM has developed neural network kernels called CMSIS-NN \cite{cmsisnn}. CMSIS-NN combined with quantization have made several mote-scale applications realizable for neural networks. Zhang et al \cite{zhang2017hello} implemented a CNN based keyword spotter, 497.6KB in size with 56.9 MOps, for a Cortex-M7. Cerutti et al \cite{cerutti2019convolutional}. implemented a 3-layer CNN for classifying 8x8 thermal images.  The SONYC-L$^3$ model is much more complex than these models.


With the VGGish model serving as the teacher network, Cerutti et al \cite{cerutti2020compact} knowledge distilled a student for a specific downstream task and implemented it on Cortex-M4. Unlike their approach, we employ embedding approximation to train an audio embedding model that can be used for transfer learning for various audio applications in New York such as scene classification. Furthermore, we chose L$^3$-Net as the teacher net over VGGish because it has fewer parameters (4.7M vs 62M) and requires less training data (296K vs 70M videos), while consistently outperforming VGGish for several downstream tasks \cite{cramer2019}.

\vspace{1mm}
\noindent
\textbf{Interference management in LoRa networks}.~~~
Self interference has been the primary focus of previous work in this area; conversely, external interference has not been dealt with systematically. Voigt et al. \cite{VoigtBRA16} for instance use multiple base stations and directional antennas for dealing with internal interference to obtain increased data extraction rates.  Zhu et al. \cite{SF-Based} and EXPLoRa \cite{ExPLoRa} allocate different spreading factors for reducing internal interference, exploiting the fact that LoRa communications with different spreading factors enjoy quasi-orthogonality due to which they do not interfere with each other. External interference is implicitly improved by range management techniques that increase the spreading factor, but doing this trades off data rate, which we avoid by using the lowest spreading factor that is compliant with single frequency FCC 15.247 access requirements.



\section{Conclusions}

Our first-of-its-kind real-time smart city monitoring system exemplifies scenarios for which it is feasible to achieve a low-power, infrastructure-agnostic, mote-scale realization of relatively sophisticated ML inferencing. Notwithstanding the considerable interest today in 5G networking (say using massive MIMO or mmWave links) for scenarios where high data rate streaming is inherent, our view is that a software-defined WSN fabric option such as ours will remain attractive for use in diverse applications, for reasons of coverage, cost, latency, or privacy.   

With an overall power consumption of 107mW, we have come close to achieving the 100mW target that we budgeted for at design time for the MKII node. We see potential to further scale SONYC-L$^3$ using compression techniques \cite{cheng2017survey}, such as parameter pruning, low-rank factorization, and efficient convolutions, to further reduce the memory requirements. 
We are also interested in extensions that support learning during field operation, using say federated learning with limited model state exchange. 

The empirical method we have developed for adaptive frequency selection deals with the key issue of external interference in a simple way.  It does, however, raise questions of learning to predict when adaption will be necessary or to detect that change is necessary. And it motivates a re-examination of whether the limitation that the discovery frequency is not adapted can be efficiently lifted.

\bibliographystyle{ACM-Reference-Format}
\bibliography{main}

\clearpage
\appendix
\section{MKII Device Details}
\vspace{1mm}
\noindent

{\bf Energy harvester analysis}.~~~
We assumed on average 3 peak sun hours per day in NYC, meaning that a 5 W harvester would produce 15 Wh per day. Three days of operation and a power conversion efficiency of 80\% would require 100 mW$\times$72 h$\times$1.25 = 9.0 Wh of energy storage. Three cells would suffice but we provisioned four cells in a series configuration for a margin of safety and to better match the 12 volt nominal output of the solar harvester. The criteria for physical size (and thus, largely, power constraints) was that the full system be small enough to be conveniently deployed by a single city worker in a busy urban setting, for example, while working out of a bucket truck.

Regarding the chosen battery chemistry, lithium-titanate (LTO) chemistry was attractive for two reasons: LTO batteries are rated to (dis)charge across a wider temperature range (in particular at the low end) and are not associated with dangerous thermal output if crushed or punctured. 
However, as sourcing LTO batteries is a potential issue, 
standard 18650 cell sizes were used 
so that an alternate chemistry could be supported with relatively minor modifications to the MKII should a different application demand it.

\vspace{1mm}
\noindent
{\bf Compute performance}.~~~The STM32H753 has a high bandwidth and relatively large capacity (2 MiB) on-die flash. These elements were key to meeting ML requirements, in particular for storing the weights. Early in development and profiling of the ML model the model weights were stored on a 100 MHz external QSPI flash chip due to their size and this setup resulted in an I/O bottleneck (with the ML inference being read intensive on the model weights) and per-iteration times in the 10s of seconds. In addition to this, the power efficiency was poor due to the CPU wasting cycles waiting for data.  Fitting the model to the much higher performance internal flash yielded per iteration times well under 1 second with performance that is almost completely CPU cycle bound rather than I/O bound at our operating point. An external 1 MiB 16-bit SRAM was added in anticipation of ML activation memory requirements; this particular need ended up being modest but the extra SRAM was ultimately crucial for other uses, such as buffering, as development evolved.

\vspace{1mm}
\noindent
{\bf Microphone selection.}~~~An effective urban noise monitoring microphone has to transduce sound consistently for long periods of time under adversarial environmental conditions. We selected the TDK InvenSense ICS-43434 \cite{ics43434}, a digital I\textsuperscript{2}S  Micro-Electro-Mechanical Systems (MEMS) microphone that has a number of favorable characteristics for this application. Within the microphone's shielded housing contains an application--specific integrated circuit (ASIC) that performs the analog to digital conversion of the analog audio signal to a digital pulse--code modulated (PCM) representation. This early stage conversion to the digital domain results in superior external radio frequency interference (RFI) and localized electromagnetic interference (EMI) rejection over purely analog designs. In addition MEMS devices were chosen for their low-cost and consistency across units, and size, which can be 10x smaller than traditional electret microphones. The microphone also has an effective dynamic range of 29--116~dBA ensuring all urban sound pressure levels can be effectively monitored.  Its frequency response was also compensated for using a digital filter as described in \cite{Mydlarz_AppliedAcoustics17}. The I\textsuperscript{2}S protocol was also convenient for low-power operation in combination with dedicated I\textsuperscript{2}S hardware in the STM32H753.

\vspace{1mm}
\noindent
{\bf SPL meter mode.}~~~MKII supports general purpose remote continuous SPL meter functionality with 32~kHz sampling and 1-second integrations with a total power consumption of 61~mW. Remote edge units report a variety of statistics such as LAeq (Level A-weighted equivalent) over the network while base stations can report each measurement. A-weighted SPL data, as specified in IEC 61672-1:2013 \cite{international2013electroacoustics}, offers a standard commonly used for noise and regulatory compliance, albeit MKII is not yet certified to any regulatory standard or type rating. However, with its favorable acoustic front-end specifications, the MKII design does have the capability to generate SPL data at the accuracy levels required for city agency noise monitoring.

\vspace{1mm}
\noindent
{\bf Enclosure design for weather tolerance.}~~~
A polycarbonate enclosure suitable for outdoor deployment from Hammond manufacturing was selected. The enclosure was modified locally with additional ports. The enclosure was weatherized to reasonably protect the contents from moisture and provide protection from insects, dirt, and other debris using by cable glands, bulkheads, etc. The result was able to withstand a simple water immersion test, not including the microphone.

\vspace{1mm}
\noindent
{\bf Clock analysis}.~~~
The STM32H753 clock tree allows us to source the acoustic sampling clock from an always-on crystal based oscillator while sourcing the CPU core clock from a PLL. Crucially, the core clock PLL can be reconfigured without disturbing the acoustic sampling, allowing for pseudo-dynamic frequency scaling to save power. Finally, a timer is sourced from an external 32.768 kHz very-high accuracy (5 ppm) low-power (order of 10 microwatts) temperature compensated oscillator. The latter is needed by the network for maintaining long-term  synchronized operation.


\begin{figure}[hbt]
 	\centering
	\includegraphics[width=0.8\linewidth]{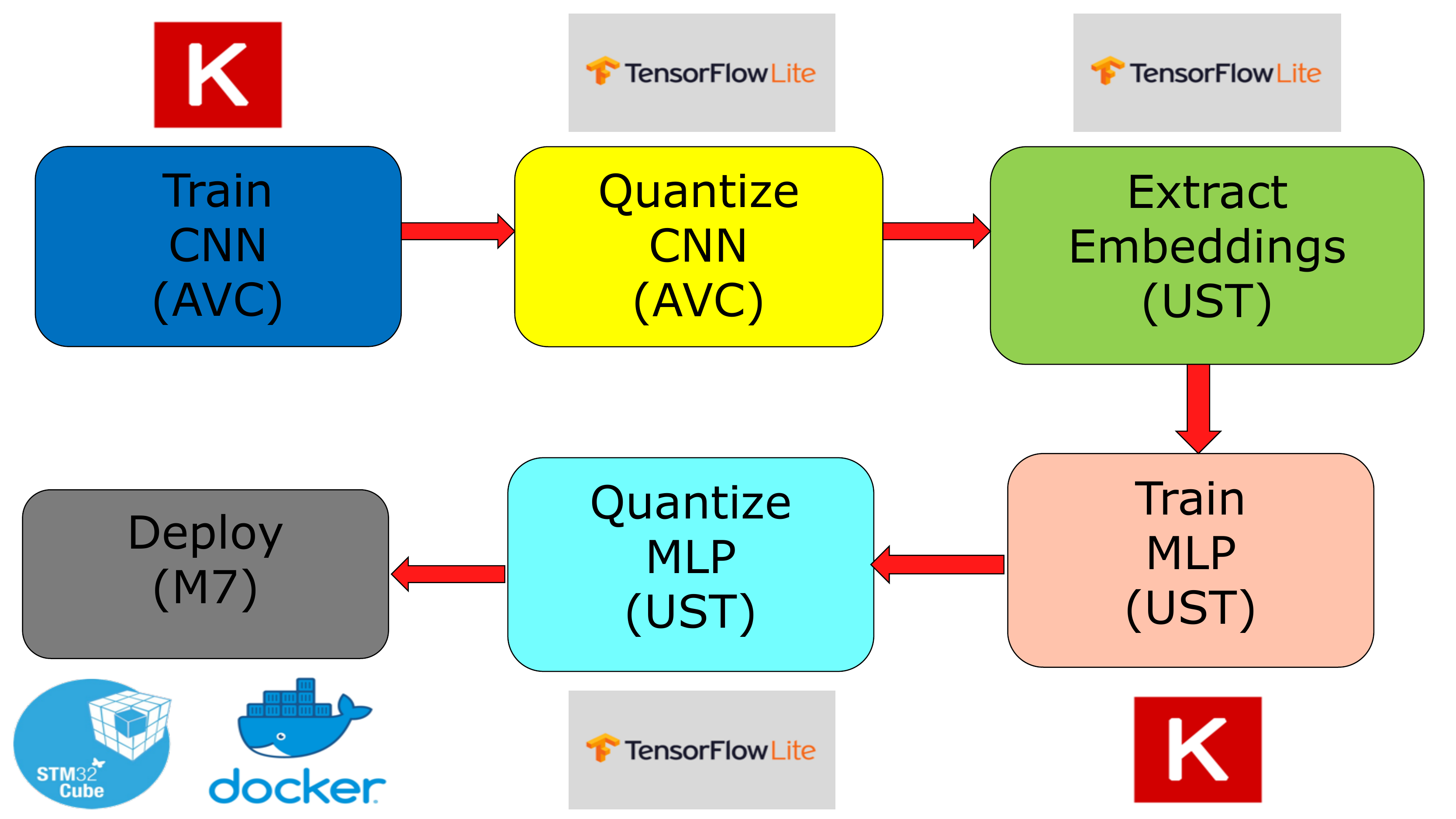}
	\vspace{-2mm}
	\caption{SONYC ML pipeline from training server to mote deployment. Associated toolchains in each step are highlighted}
	\label{fig:sonyc_pipeline}
\end{figure}

\vspace{1mm}
\noindent
{\bf Programmability Support}.~~~A key efficiency in the development process was establishing a semi-automated model deployment pipeline from the training server to the MKII mote (Figure \ref{fig:sonyc_pipeline}). Its key steps and the toolchains associated with each step are as follows: (i) Training SONYC-L\textsuperscript{3} on the semi-supervised AVC task using Keras, (ii) Quantizing the featurizer, once again on the semi-supervised AVC task, using TensorFlow Lite, (iii) Training and quantizing the end classifier using Keras and TensorFlow Lite, (iv) Translating the quantized models into native code for the Cortex-M7 using STM32 X-CUBE-AI \cite{xcube}, (v) Model integration into the eMote \cite{emote} OS running on the edge devices. eMote is a hybrid platform that allows for programming in high-level languages (e.g., C\#) where possible, but also directly integrating performance/power critical code close-to-the-metal (with minimal abstraction).

While the aforementioned pipeline in its current form is semi-automated and requires a human in the loop, it can be fully automated for continuous integration and continuous delivery (CI/CD) and we intend to support this in a future release.  Changes to the model or application can be deployed directly on a base station wired to a MKI gateway through a Docker pipeline. Remote model updates to edge devices are currently not supported due to the bandwidth limitations on the LoRa network.



\section{MKII Link Details}

\begin{figure}[hbt]
	\centering
		\begin{subfigure}[h]{0.47\textwidth}
			\centering
            \includegraphics[width=\textwidth, trim = 20 0 25 5, clip]{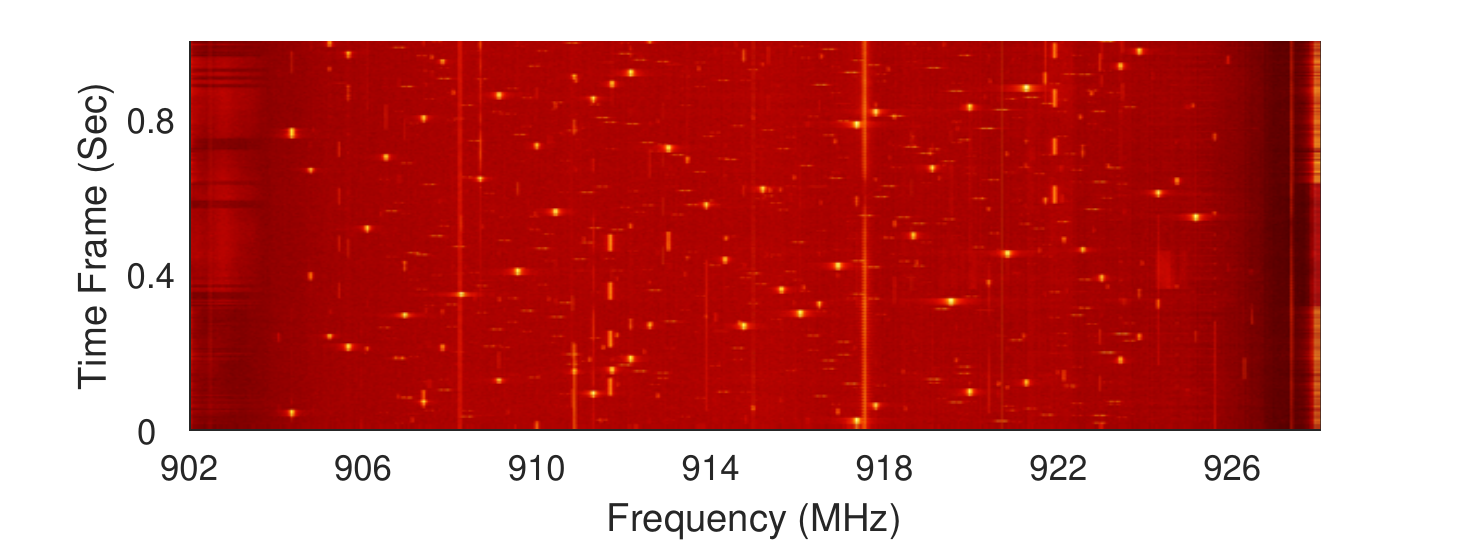}
			\caption{Overlooking Times Square}
		\end{subfigure}%
		
		\begin{subfigure}[h]{0.47\textwidth}
			\centering
			\includegraphics[width=\textwidth, trim = 20 0 25 5, clip]{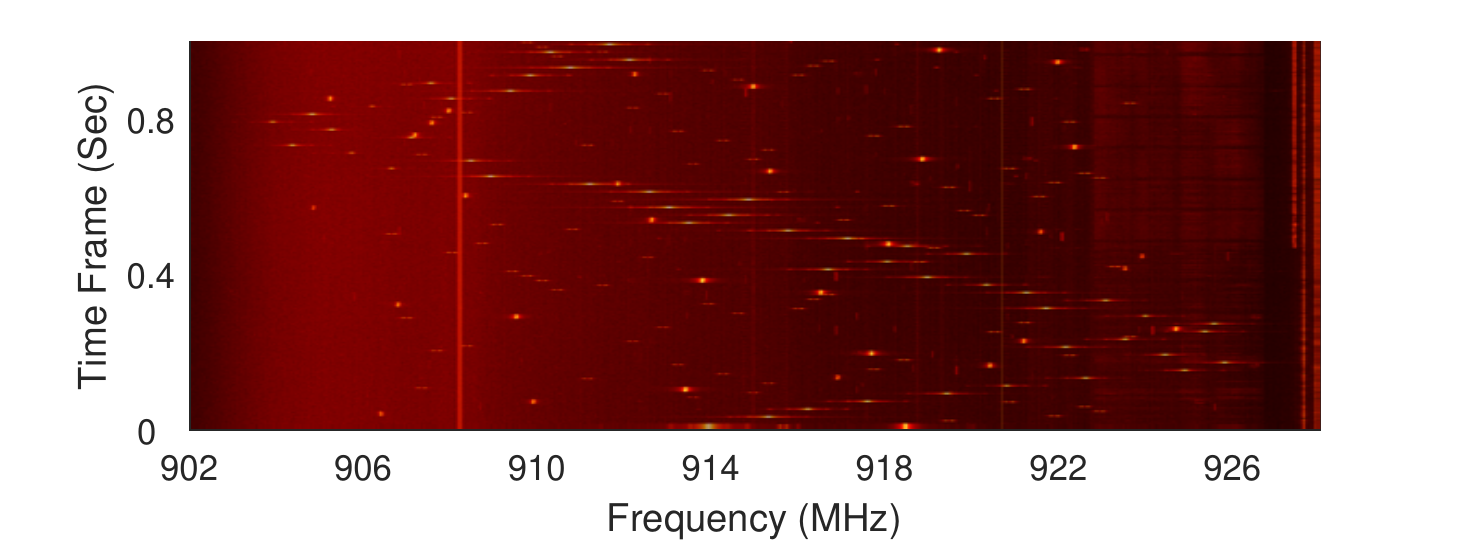}
			\caption{
			On Jay Street in Brooklyn}
		\end{subfigure}%
		\vspace{-2mm}
	\caption{902-928 MHz ISM band frequency spectrograms collected using a LimeSDR \cite{limesdr} at representative NYC locations showing significant external interference of different sorts}
	\label{network_Spec_results}
\end{figure}

Figure \ref{network_Spec_results} corroborates the observation in Section 5 that external interference is commonly experienced by links operating in different locations and times. Interference is apparent in these visualizations even though are only over 1 second snippets chosen randomly from longer collects.  Brighter (resp., darker) colors in the spectrogram correspond to stronger (resp., weaker) external interference.  Unsurprisingly, external interferers are of different sorts (i.e., narrowband vs.~wideband, fixed frequency vs.~frequency-hopping), and vary with location and frequency.  

\begin{figure}[hb]
	\centering
	\includegraphics[width=0.47\textwidth, trim = 20 0 25 5, clip]{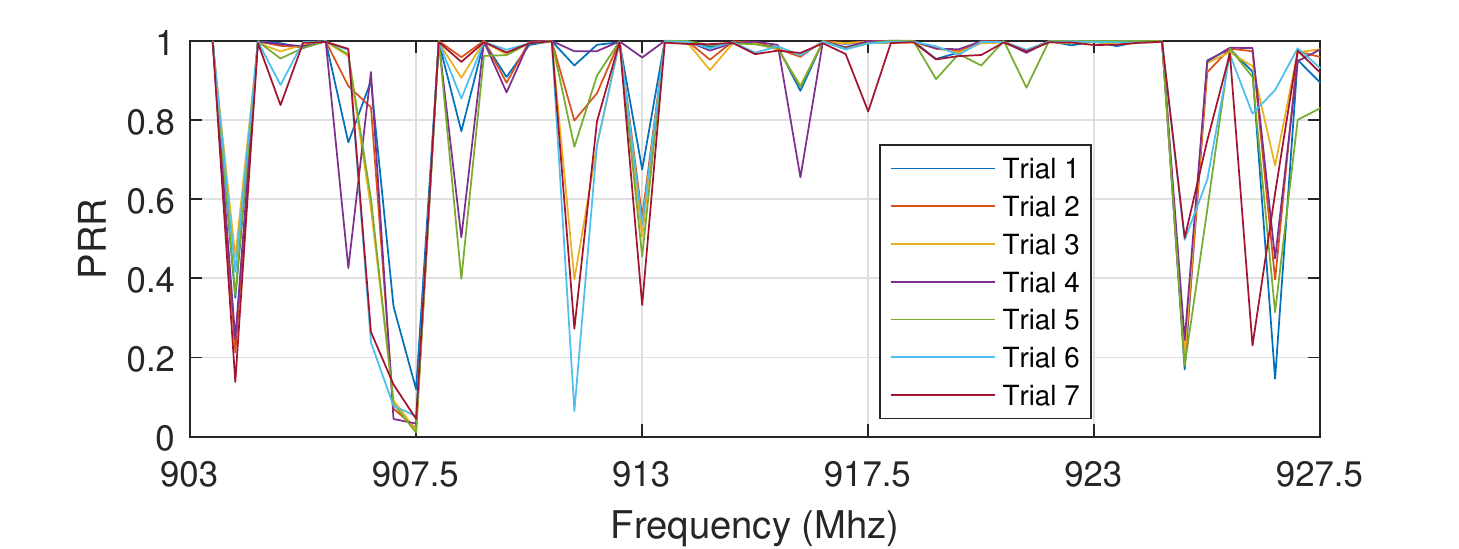}
	\vspace{-2mm}
	\caption{Across day long trials in our Tier 2 deployment, some frequencies have consistently high PRR over the link between a pair of MKII nodes}
	\label{network_LoRa_atob_results}
\end{figure}

\begin{figure}[hb]
	\centering
	\includegraphics[width=0.47\textwidth, trim = 20 0 25 5, clip] {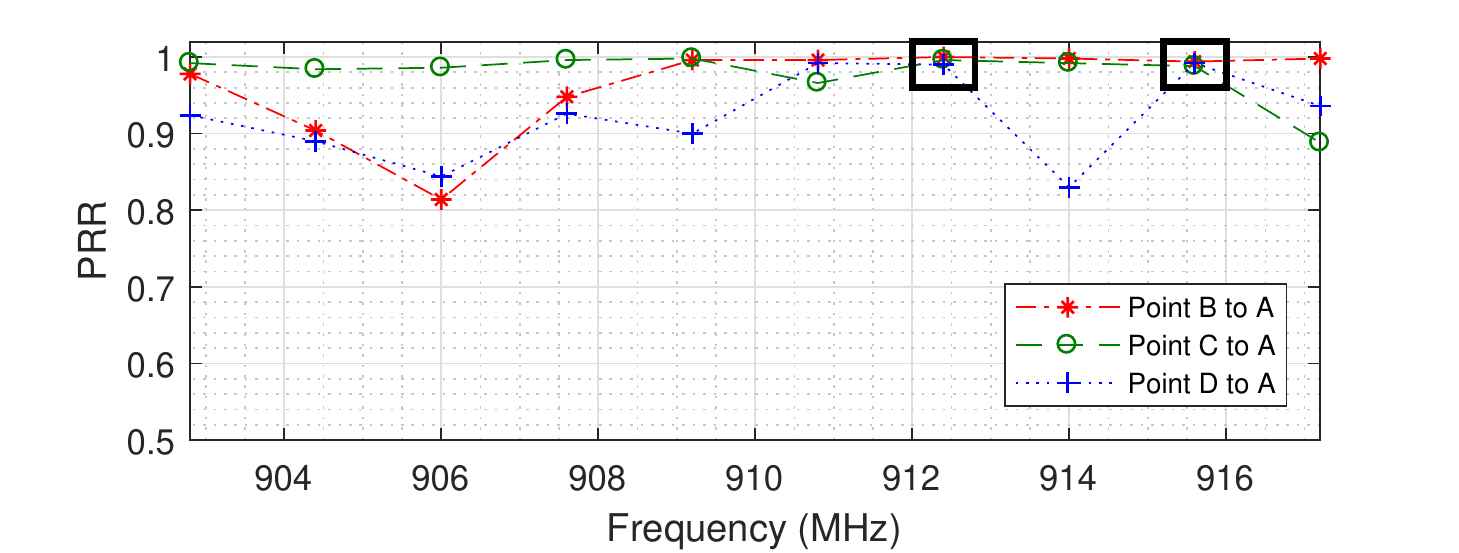}
	\vspace{-2mm}
	\caption{Packet reliability across links in a network, deployed in a different city, to a common destination MK II node named A, have some common channels with consistently high PRR}
	\label{network_goodchannel}
\end{figure}

Figures \ref{network_LoRa_atob_results} and \ref{network_goodchannel} exemplify that even though PRR varies with frequency, possibly due to variation in external interference and frequency selective or shadowing related fading, in each network/location there exist frequencies that yield high PRR. 

\end{document}